\documentclass{article}
\usepackage[english]{babel}
\usepackage{amsfonts}
\usepackage{amssymb}
\usepackage[dvips]{graphicx}
\usepackage{subfigure}
\usepackage{color}

\begin{document}

\title{ Experimental study of out of equilibrium
fluctuations in a {colloidal suspension of Laponite}
using optical traps}
\author{Pierre Jop, J. Ruben Gomez-Solano, Artyom Petrosyan and Sergio
Ciliberto $^{\ast}$ \\
$^{\ast}$ Corresponding author.
Email:sergio.ciliberto@ens-lyon.fr \\
Universit\'e de Lyon, Laboratoire de Physique,\\ Ecole
Normale Sup\'erieure de  Lyon, CNRS ,\\
        46, All\'ee d'Italie, 69364 Lyon CEDEX 07, France}
\maketitle

\begin{abstract}

This work is devoted to the study of displacement fluctuations of
micron-sized particles in  an aging colloidal glass. We address the
issue of the validity of the fluctuation dissipation theorem (FDT)
and the time evolution of viscoelastic properties during aging of
aqueous suspensions of a clay (Laponite RG) in a colloidal glass
phase. Given the conflicting results reported in the literature for
different experimental techniques, our goal is to check and
reconcile them using \emph{simultaneously} passive and active
microrheology techniques. For this purpose we measure the thermal
fluctuations of micro-sized brownian particles immersed in the
colloidal glass and trapped by optical tweezers. We find that both
microrheology techniques lead to compatible results even at low
frequencies and no violation of FDT is observed.
 Several interesting features concerning the statistical properties
 and the long time correlations of the particles are observed during the transition.
\end{abstract}

\section{Introduction}

The description of out-of-equilibrium systems is a topic of major
interest in physics since in most systems, energy or matter flows
are not negligible but drive them into non-ergodic states for
which equilibrium statistical mechanics may  be not applicable.
The typical examples of such systems are slowly stirred systems
and glasses. Unlike equilibrium statistical mechanics, whose
foundations are well established since long time ago,
non-equilibrium statistical mechanics is still being developed.

One attempt to develop a statistical mechanics description of
non-equilibrium slowly evolving systems is to extend some
equilibrium concepts to them, namely the so called fluctuation
dissipation theorem (FDT). FDT relates the power spectral density of
equilibrium fluctuations { of a given observable} $x$ of systems in
contact with a thermal bath at temperature $T$ to the response
$\chi$ to a \emph{weak} external perturbation. In frequency domain
the relation between the corresponding Fourier transforms involves a
prefactor given by $T$
\begin{equation}\label{}
    \langle |\hat{x}(\omega)|^2 \rangle = \frac{4 k_B T}{\omega} Im\{
    \hat{\chi}(\omega)\}.
\end{equation}
Despite of the fact that a thermodynamic temperature is not
rigorously defined for non-equilibrium systems, a generalization
of FDT can be done by means of an \emph{effective temperature}
$T_{eff} (\omega, t_w)$ which describes fluctuations at a time
scale $1/\omega$. It is defined as
\footnote{{It is worth noting that our definition of $T_{eff}$ differs
slightly from the one proposed by Pottier \cite{pottier}, but its behavior
is qualitatively the same.}}:
\begin{equation}\label{eq:Teff}
    T_{eff}(\omega, t_w) = \frac{\omega \langle |\hat{x}(\omega, t_w)|^2 \rangle} {4 k_B
    Im\{
    \hat{\chi}(\omega, t_w)\}},
\end{equation}
where $t_w$ denotes the \emph{age} of the system, i.e. the waiting
time since the systems left equilibrium. In general, for a
non-equilibrium system either the spectral density of fluctuations
and the response function depend on $t_w$ and then violations of
FDT  ($T_{eff}(\omega,t_w)\neq T$) could also depend on it.

Glasses are one of the examples of non-equilibrium systems that have
been studied extensively over the past years. They are disordered
systems at microscopic length scales formed after a quench from an
ergodic phase (e.g. an ordinary liquid in equilibrium) to a
non-ergodic phase (e.g. a supercooled liquid). After a quench, a
glassy system relaxes asymptotically to a new phase in equilibrium
with the new temperature $T$ of its surroundings, but the time
needed to reach equilibrium may be {extremely large}. Since the
system is non-ergodic during its relaxation, there is no invariance
under time translations and then its \emph{age} $t_w$ is well
defined. The slow time evolution of their physical properties (e.g.
viscosity) is known as \emph{aging}. A number of theoretical,
numerical and experimental studies have shown that two different
effective temperatures are found for structural glasses
\cite{parisi}, \cite{kob} and spin glasses \cite{cugliandolo},
\cite{fielding}, \cite{herisson}. One of them is the effective
temperature associated to the fast rattling fluctuations (high
frequency modes) and is equal to $T$. The other effective
temperature is the one associated to the slowest structural
rearrangements (low frequency modes) whose value is larger than $T$
and decreases with time to $T$. Both effective temperatures have the
properties of thermodynamic temperatures in the sense that they
correspond to two different thermalization processes at different
time scales \cite{kurchan}.

\subsection{FDT in Laponite suspensions}

Regarding the existence of an effective temperature higher than
the bath temperature for structural and spin glasses, several
recent works have attempted to look for violations of FDT in
another kind of non-equilibrium systems: \emph{colloidal glasses}.
Unlike structural or spin glasses, the formation of either  a
colloidal glass or a gel \cite{bonn_PRE_2004} does not require a
temperature quench but the packing of colloidal particles at a
certain low concentration in water forming a glassy structure.

Aqueous suspensions of clay \emph{Laponite} have been studied as a
prototype of colloidal glasses. Laponite is a synthetic clay formed
by electrostatically charged disc-shaped particles of chemical
formula
Na$^+_{0.7}$[Si$_8$Mg$_{5.5}$Li$_{0.3}$O$_{20}$(OH)$_4$]$^{0.7-}$
whose dimensions are 1 nm (width) and 25 nm (diameter). When
Laponite powder is mixed in water, the resulting suspension
undergoes a sol-gel transition in a finite time (e.g. a few hours
for 3 wt \%), i.e. it turns from a viscous liquid phase (sol) into a
viscoelastic solid-like phase (gel). During this aging process,
because of electrostatic attraction and repulsion, Laponite
particles {form a \emph{house of cards}-like structure.}

Several properties of Laponite suspensions during aging have been
extensively studied, such as viscoelasticity \cite{abou0},
translational and rotational diffusion \cite{jabbari} and optical
susceptibility \cite{ruocco} using techniques such as rheology and
dynamic light scattering. However, the available results obtained so
far for the issue of the validity of the FDT in this system are
contradictory. Bellon \emph{et al}
\cite{Bellon02,Bellon02_I,Bellon02_II} reported an effective
temperature from dielectric measurements indicating a strong
violation of FDT at low frequencies ($f < 40$ Hz), whereas the same
group did not observe any violations from mechanical measurements
\cite{Bellon02_I}. The reason of this difference between mechanical
and dielectric  measurements, which is not yet very well understood,
has been already widely discussed \cite{Bellon02_I} so we will not
insist on this point. We can only mention that in principle there
are not deep reasons that the $T_{eff}$ obtained from FDR using
different observables has to be the same. Furthermore dielectric
measurements are strongly affected by the ions inside the solution
which may be the source of the violation of FDT \cite{Bellon02_I}.
The interest here is on the mechanical measurements. The rheological
measurement described in { Ref.} \cite{Bellon02_I} was a global
measurement and one may wonder whether an experiment of
microrheology give or not the same results. This experiment can be
performed using as a probe a Brownian particle using active and
passive microrheology. In these kind of experiments, the issue is to
measure whether the properties of the Brownian motion are affected
by the fact that the surrounding fluid (the thermal bath) is out of
equilibrium. Several experiments of Brownian motion inside a
Laponite solution have been done by different groups using various
techniques. The results are rather contradictory. Let us {
summarize} them. Abou \emph{et al} \cite{Abou03} observed that
$T_{eff}$ increases in time from the value of the bath temperature
to a maximum and then it decreases to the bath temperature.
Jabbari-Farouji \emph{et al} \cite{jabbari1} used a combination of
\emph{passive} and \emph{active microrheology} techniques (see
Subsection \ref{microrheo}) without any observation of deviations of
the effective temperature from the bath temperature over several
decades in frequency (1 Hz$-$10 kHz). Greinert \emph{et al}
\cite{greinert} observed that the effective temperature increases in
time using a passive microrheology technique. Finally,
Jabbari-Farouji \emph{et al} \cite{jabbari2} found again no
violation of FDT for the {frequency range of 0.1 Hz$-$10 kHz}.

In view of the conflicting results obtained for different
experimental techniques we have compared them in order to understand
where the difference may come from. The purpose of this article is
to describe the results of the measurement of the Brownian motion of
either one or several particles inside a Laponite solution using
optical tweezers manipulation and a combination of different
techniques proposed in previous references. The measurements
performed with several particles are of particular importance
because they allow us to compare simultaneously passive and active
microrheology techniques (see { Sect. 3} for a definition) in the
same Laponite suspension. Some preliminary results have been already
described \cite{jop} and we extend the results in this article.

All the techniques do not show, within experimental errors, any
 increase of $T_{eff}$ which  remains equal to that of the
 thermal bath for all the duration of the sol-gel transition.
Thus the result of this papers agrees with those of
Jabbari-Farouji et al. \cite{jabbari1,jabbari2} and of Bellon et
al. \cite{Bellon02_I}. A tentative explanation of the discrepancy
of the results is also given and we show that may be induced by
the data analysis.
 The article is organized as { follows}. In { Section 2} we describe the
 experimental apparatus. In { Section 3} the various techniques used to measure
 response and fluctuations are introduced.  In { Section 4} we describe the
 experimental results of the single bead experiment. In { Section 5}
 the results on the multiple beads technique are reported. Finally in { Section 6}
 we discuss the results and we conclude.

\section{Experimental set up}

\subsection{Sample preparation} \label{section_sample_preparation}

\begin{figure*}
  \centering
  \includegraphics[width=10.0cm,height=6.0cm]{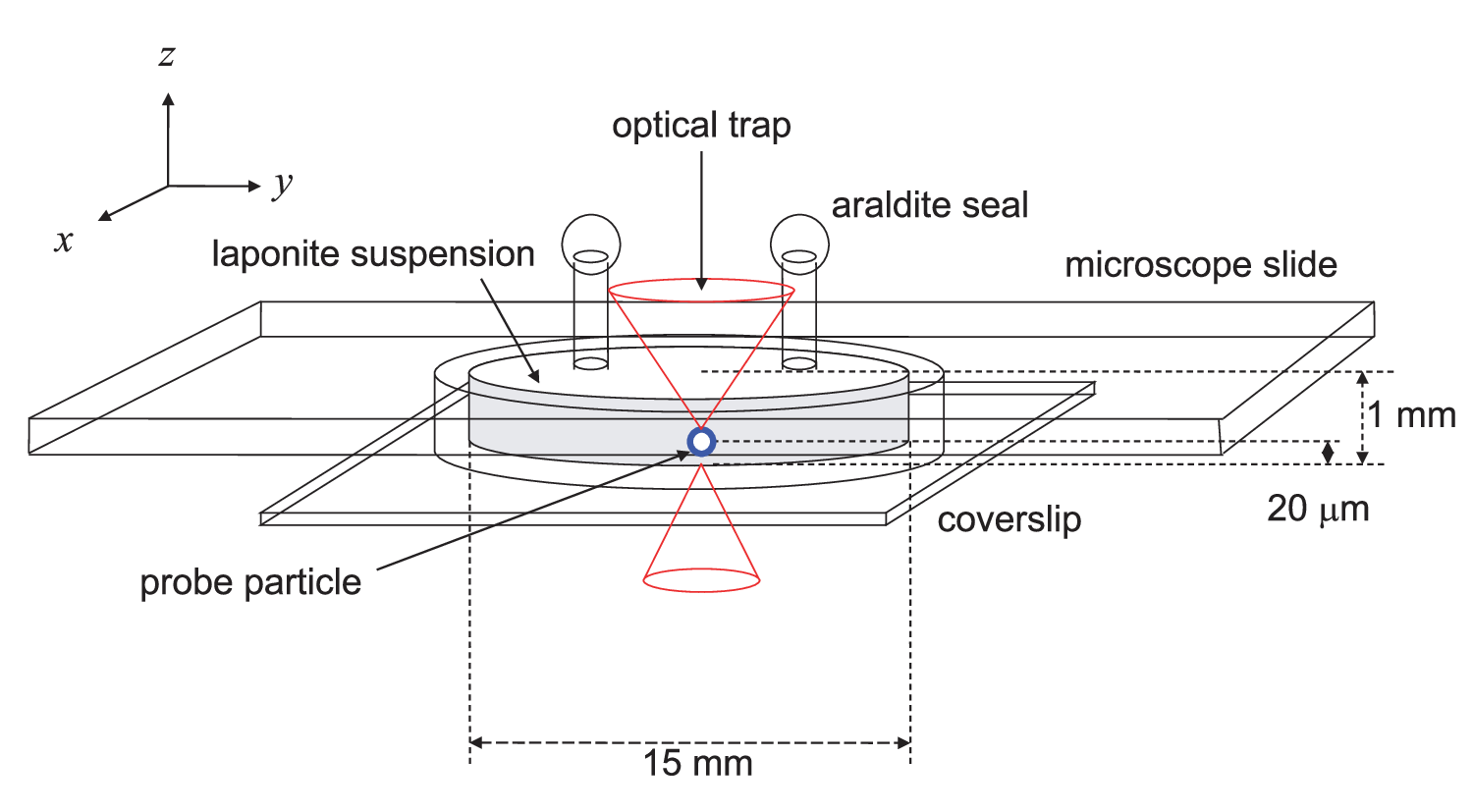}
  \caption{Diagram of the sample cell used during the experiment.
  The diagram is not in proper scale.}
  \label{fig:samplecell}
\end{figure*}

Physical properties of Laponite suspensions are very sensitive to
the method used during their preparation \cite{cummins}. Hence, an
experimental protocol must be followed in order to perform
reproducible measurements of their aging properties. Laponite RD,
the most frequently studied grade and the one studied in this work,
is a hygroscopic powder that must be handled in a controlled dry
atmosphere. The powder is mixed with ultrapure water, at a weight
concentration which has been varied from 1.2 to { 3.0} \%. The pH
has been adjusted to 10 in order to assure chemical stability of the
samples. At lower pH, the decomposition of clay particles occur.
CO$_2$ absorption by water can modify the pH of the samples and then
their aging properties. For these purposes, the preparation of the
samples is done entirely within a glove box filled with circulating
nitrogen. As we said the concentration { is} varied from 1.2 \% to {
3.0}  \% wt for different ionic strength. These conditions allow us
to obtain either a gel or a glass according to the phase diagramm
found in the literature \cite{bonn_PRE_2004}. At 2.8 \%  Laponite
gelation takes place in few hours after the preparation. The
suspension is vigorously stirred by a magnetic stirrer during 30
minutes. The resulting aqueous suspension is filtered through a 0.45
$\mu$m micropore filter in order to destroy large particle
aggregates and obtain a reproducible initial state. The initial
aging time ($t_w = 0$) is taken at this step. Immediately after
filtration, a small volume fraction (0.02 { $\mu$l} per 50 ml of
sample) of silica microspheres (diameter = 2 $\mu$m) is injected
into the suspensions. Microspheres play the role of probes to study
the aging of the colloidal glass, as explained further. The sample
is placed in a ultrasonic bath during 10 minutes to destroy small
undesired bubbles that could be present during the optical tweezers
manipulation. The suspension is introduced in a sample cell which is
sealed in order to avoid evaporation and the use of vacuum grease,
used in other experiments. Indeed this grease, whose pH is much
smaller than 10, quickens the evolution of the suspension giving
very strange and not reproducible results. Two kinds of completely
sealed cells have been used. The first used one microscopic slide
and a coverslip glued with Gene Frame adhesive spacers. This cell
gives good results but the contact surface between the suspension
and the adhesive is large. In order to decrease such contact surface
we use more recently a sample cell consisting on a microscope slide
and a coverslip separated by a cylindrical plastic spacer of inner
diameter 15 mm, thickness 1 mm and glued together with photopolymer
adhesive, as shown in Fig.~\ref{fig:samplecell}. In this cell the
inputs are sealed with araldite adhesive in order to avoid
evaporation and direct contact with CO$_2$ from air.

\subsection{Optical tweezers setup and data acquisition}

We measure the fluctuations of the position of one or several silica
beads trapped by an optical tweezers during the aging of the
Laponite. The experiment is performed in a typical optical tweezers
system  at room temperature ($22 \pm 1^{\circ}$ C).

\subsubsection{Single Trap}

The laser beam ($\lambda=$ 980 nm) is focused by a microscope
objective ($\times$63) 20 $\mu$m above
 the coverslip surface to create a
 harmonic potential well where a bead of 1 or 2 $\mu$m in diameter ($2r$) is trapped.
 The power of the trapping beam is varied from 33 mW to 71 mW. With these laser
 intensities
 the  surrounding liquid is heated of only a few degree.
 The position of the bead is detected by measuring with a 4 quadrant detector,
 the deflection of a He-Ne laser beam
  which is focused on the bead by the same microscope objective.
 The power of the measuring beam is  only a few mW.
 The output signals of the 4 quadrant detector, which measures the positions { $(x,y)$} of the bead, are acquired at a sampling frequency of {
8196 Hz} with {16-bit} accuracy.

\begin{figure}
 \centering        {\includegraphics[width=.39\textwidth]{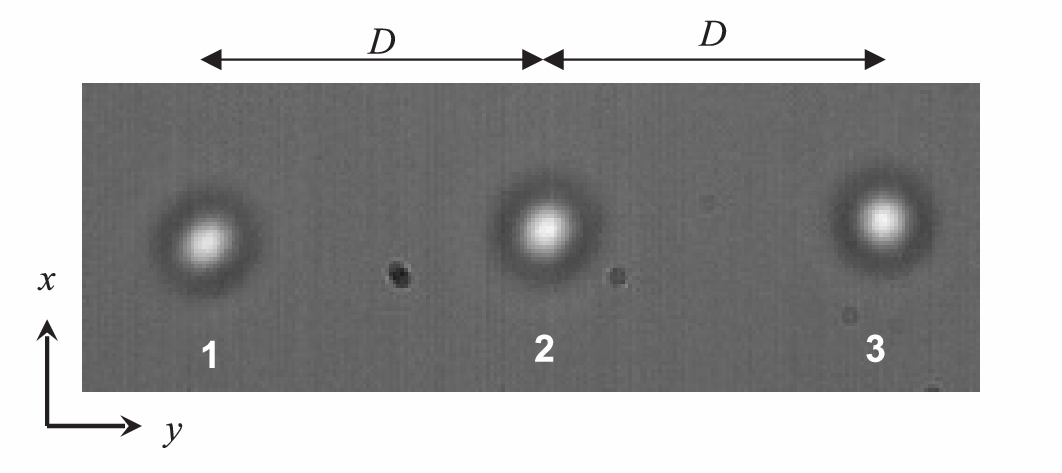}}
     \caption{Configuration of three optical traps separated by a distance
     $D = 9.3 \times 10^{-6}$ m. The bright spots in the image correspond
     to three probe particles of 2 $\mu$m diameter trapped by them.}
  \label{fig:threetraps}
\end{figure}

\subsubsection{Multiple Trap}

 We use a Nd:YAG diode pumped
solid state laser (Laser Quantum $\lambda = 1064$ nm) whose beam is
strongly focused by a microscope oil-immersion objective ($\times
63$, NA $=1.4$). The sample cell is placed on a piezoelectric stage
(NanoMax-TS MAX313/M) in order to have mechanical control in three
dimensions of nanometric accuracy during the trapping process. The
sample cell is placed with the glass plate at the top and the
coverslip at the bottom in such a way that the focus of the laser
beam is located 20 $\mu$m above the coverslip inner surface (see
Fig.~\ref{fig:samplecell}). The whole system is mounted on an
optical table in order to get rid of low frequency mechanical noise.

The light source used for detection is a halogen lamp whose beam
is focused in the sample by a condenser lens. The halogen light
passes through the sample and then the resulting interference
contrast signal is detected by a high speed camera (Mikrotron
MC1310) after emerging from the microscope objective. Once probe
particles are trapped properly, images with a resolution of
240$\times$80 pixels are recorded at a sampling rate of 200 frames
per second during 50 seconds and saved in AVI format every 97
seconds (i.e. 10000 images per AVI file) for further data
processing. Hence, the aging time of the system $t_w$ is measured
in multiples of 97 s while the smallest measurable time scale (the
time elapsed between successive positions of the probe)
corresponds to 5 ms and the highest accessible frequency is 100
Hz. { We follow the trajectories of the trapped bead
($(x(t),y(t))$ on the focal plane
by means of a MATLAB image processing program.}

Multiple probe particles are needed to be trapped within the same
sample in order to perform simultaneously passive and active
microrheology measurements (see { Sect. \ref{microrheo}}). Since two
trapped particles are needed for each technique, at least three
different optical traps must be created simultaneously, as shown in
{ Fig. \ref{fig:threetraps}}. For passive measurements, two traps of
different stiffness and fixed positions are needed. We label the
weakest trap as '1' while the strongest one as '2'. For active
measurements, one needs one trap in a fixed position in order to
measure the power spectral density of displacement fluctuations. For
convenience, we choose trap '2' for such purpose. In addition, one
needs an oscillating trap in order to measure the response of the
probe to an external forcing. This trap is labeled as '3'. The
separation distance between adjacent traps is $D = 9.3 \times
10^{-6}$ m, which is sufficiently large to avoid correlations
between their motions. { In the following, $x$ defines the direction
along which the position of the optical trap is oscillated in time,
while $y$ corresponds the orthogonal direction (see Fig.
\ref{fig:threetraps})}. For simplicity, all our calculations rely on
$x$ coordinates only.

The creation of three optical traps on the $x-y$ plane is
implemented by means of two coupled acousto-optic deflectors (AOD).
In order to create three traps, the laser beam scans three different
positions along the $y$ direction { at a rate of 10 kHz}. The
scanning rate must be large enough in order to avoid the diffusive
motion of the particle through the surrounding medium during the
absence of the beam. The oscillation of the position of trap '3' is
accomplished by deflecting the laser beam along $x$ with a given
waveform $x_0(t)$ when it scans the position that it would visit in
the absence of deflection along $x$. The stiffness of each trap is
proportional to the time that the laser beam stays in the
corresponding position. We check that by selecting a ratio of
20:40:40 for the visiting time of traps '1', '2' and '3',
respectively we obtain a stiffness ratio of { 20.7:39.7:39.6,
respectively. Specifically the stiffness of the three traps are:
$k_1 = 3.73 \: \textnormal{pN}/\mu \textnormal{m}$, $k_2 = 7.15 \:
\textnormal{pN}/\mu \textnormal{m}$, $k_3 = 7.12 \:
\textnormal{pN}/\mu \textnormal{m}$. The stiffness has been measured
using the power spectrum technique of Ref. \cite{capitanio}.}

\section{Microrheology techniques}\label{microrheo}

The determination of effective temperature, viscosity and elasticity
of laponite samples during aging is done using microrheology. It
consists on the measurement of the motion of probe particles
embedded in the colloidal glass and manipulated by the optical
tweezers system. This is the approach followed by
\cite{jabbari1,greinert,jabbari2} leading to conflicting results
between passive and active measurements. There are two different
kinds of microrheology techniques depending on the manipulation of
the probes by the optical trap: \emph{passive} and \emph{active}.

In order to measure $T_{eff}$ for the particle motion several
techniques have been used. The first one is based on the laser
modulation technique as proposed in \cite{greinert}. The second is
based on the Kramers-Kronig relations with two laser intensities and
it combines the advantages of two methods
proposed in \cite{KK_R,jabbari} and in \cite{greinert}.
Finally the third  method uses the advantages of several beads in
a multi-traps tweezers and allows to apply the different
techniques simultaneously.

\subsection{Passive microrheology}

{ In passive microrheology (PMR), an optical trap acts as a passive
element keeping a bead in a fixed position
while measuring the displacement fluctuations $x(t)$
to probe the properties of the medium.}

In the present work, PMR measurements of effective temperature and
elasticity of the colloidal glass are based on the method proposed
by Greinert \emph{et al} \cite{greinert} This method is based on a
generalization of the equipartition relation to glassy systems
argued by Berthier and Barrat \cite{berthier}. Since a Laponite
suspension becomes viscoelastic as it ages, a probe particle
immersed in it is subject to an additional force $F_e =- K_{Lap}
x$ due to the elasticity of the colloidal glass, where $K_{Lap}$
denotes the effective elastic stiffness of the colloid. Then, by
introducing an effective temperature $T_{eff}$ by means of the
generalized equipartition relation
\begin{equation}
  { k_B \ T_{eff} \over (k+K_{Lap})} = \left<
  x^2-\left<x\right>^2\right>,
   \label{eq:equi}
\end{equation}
where brackets stand for average over the time (during a time window
such that $K_{Lap}$ is almost constant). Note that this definition
of effective temperature is global in frequency and then its
measurement is strongly limited by the capability of detecting slow
modes.

\subsubsection{Laser intensity modulation technique}

Following the method used in \cite{greinert}, the stiffness of the
{single} optical trap is periodically switched between two different values
($k_1$ and $k_2$) every 61 seconds by changing the laser
intensity.
At a given $t_w$, on an interval of { $60$ s}  we compute the
variance of the position $ \langle \delta x^2(t_w) \rangle =\langle
x^2-\left<x\right>^2\rangle$ on time windows of length { 3 s
$<\tau<$ 50 s}. One must be aware that this time window is not
sufficiently long to take into account the role of low frequency
fluctuations at late stages during aging, and variances could be
underestimated in this regime, as noted by \cite{jop} and in { Sect.
4}.  The values of the variance computed in each window are then
averaged  over
 a 60 s interval, which is  chosen sufficiently short to
assure that the viscoelasticity of the colloidal glass remains
almost constant, specifically :
\begin{equation}
\langle \delta x^2 (t_w) \rangle_j= {1 \over \tau
t_{max}}\int_{t_w}^{t_w+t_{max}}dt \int_t^{t+\tau}
(x_j^2(t')-\langle x_j(t') \rangle^2) dt',
\end{equation}
where { $t_{max}=60$} s and $\langle \delta x_j^2 (t_w) \rangle$
with $j=1,2$ correspond to the variance measured with $k_1$ and
$k_2$ respectively. To avoid transient each record is started 20 s
after the laser switch to be sure that the system has relaxed toward
a quasi stationary state. Assuming the equipartition principle
Eq.~(\ref{eq:equi}) still holds
  in this out-of-equilibrium system, $T_{eff}$
 is computed as in \cite{greinert}.
 The expressions of the effective temperature
 and of the Laponite elastic stiffness $K_{Lap}$
  are the following
\begin{equation}
   k_BT_{eff}={
   (k_2 - k_1) \langle \delta x^2(t_w) \rangle_1 \langle \delta x^2(t_w) \rangle_2
   \over \langle \delta x^2(t_w) \rangle_1 - \langle \delta x^2(t_w)
   \rangle_2},
\label{eq:temperature}
\end{equation}
\begin{equation}
   K_{Lap}=
   {k_1\ \langle \delta x^2(t_w) \rangle_1- k_2 \langle \delta x^2(t_w) \rangle_2
   \over
  \langle \delta x^2(t_w) \rangle_1-\langle \delta x^2(t_w)
  \rangle_2}.
\label{eq:KLap}
\end{equation}

We notice that when multiple traps are used there is no need to
modulate the laser as the stiffness of the trap  '1' and '2' in Fig.
\ref{fig:threetraps} are respectively  $k_1$ and $k_2$. Thus the
method is immediately implemented using these two beads. The only
difference is that now the acquisition frequency is 200 Hz. But
there are several advantages here. The two variances at $k_1$ and
$k_2$ are computed simultaneously and {since the power of the laser
is not commuted there are neither temperature variation inside the
sample nor dead time of} { 20 s}. Thus the measurement of the
variance can be done each { 60 s}. In this way the results  are
cleaner than with a single bead.

This technique, although quite interesting and simple,  has several
drawbacks that we will discuss in the next sections. Furthermore,
being a global measurement, it has no control of what is going on on
the different frequencies. To overcome this problem we have
implemented the following {method, which can be used
only with a single bead due to the need of} { a}
{ large sampling frequency.}

\subsubsection{Kramers-Kronig and modulation technique}

This combines  the laser modulation technique described in the
previous section and the passive rheology technique based on
Kramers-Kronig relations. The fluctuation dissipation relations
relate the spectrum $S_j(\omega)$ of the fluctuation of the
particle position to the imaginary part $\chi_j''$ of the response
of the particle to an external force, specifically :
\begin{equation}
S_j(\omega,t_w)= {4 \ k_B T_{eff} \over \omega} \chi_j''(\omega,t_w)
\label{eq:FDR},
\end{equation}
with $j=1$ and $j=2$ for the spectra measured with the trap
stiffness $k_1$ and $k_2$  respectively. We recall that for a
particle inside a  newtonian fluid $G=1/\chi_j= k_j+ i\gamma \omega$
with $\gamma=6\pi\eta r$ the drag coefficient and $\eta$ the fluid
viscosity.
In Eq.~(\ref{eq:FDR}) the dependence in $t_w$ takes into account the
fact that the properties of the fluid changes after the preparation
of the Laponite suspension. If one assumes that $T_{eff}$ is
constant as a function of frequency (hypothesis that can be easily
checked a posteriori) then the real part $\chi_j'$ of the response
is related to $\chi_j''$ by the Kramers-Kronig relations \cite{KK_R}
that is:
\begin{equation}
\hat \chi_j'(\omega,t_w)={ 2\over \pi } P\int_0^\infty {\xi \hat
\chi''(\xi,t_w)  \over \xi^2-\omega^2} d\xi ={ 1\over 2 \pi k_B T }
P\int_0^\infty {\xi^2 S_j(\xi,t_w)  \over \xi^2-\omega^2} d\xi
\label{eq:KrKr},
\end{equation}
where $P$ stands for principal part of the integral and $\hat
\chi_j'$ is  computed assuming that $T_{eff}$ does not depend on the
frequency in the second equality  where we have used
Eq.~(\ref{eq:FDR}), i.e. $\hat \chi''(\xi,t_w)= \omega
S_j(\omega,tw) / (4k_BT) $. To compute $\hat \chi_j'$ from
Eq.~({\ref{eq:KrKr}}) we use a Fourier transform algorithm that is:
\begin{equation}
\hat \chi_j'(\omega,t_w)= { 1\over 2 \pi k_B T }
\int_0^{1/\omega_{min}} \cos(\omega \ t) \ dt \
\int_0^{\omega_{max}} \xi^2 S_j(\xi,t_w)
 \ \sin(\xi \ t) d\xi,
\label{eq:KrKr_F}
\end{equation}
where $\omega_{min}$, $\omega_{max}$ are the minimum and maximum
frequencies  resolved in $S_j(\omega)$. Mathematically one should
have  $\omega_{min}=0,\omega_{max}=\infty $. However this is not
possible experimentally and the limited interval
$[\omega_{min},\omega_{max}]$ induced an error in the high frequency
estimation of $\hat \chi_j'$. This error can be easily estimated and
one evaluates that the values of $\chi'$ are correct within a few
percent  for $ \omega \le \omega_{max}/10$. Once $\hat \chi_j'$ has
been computed from Eq.~(\ref{eq:KrKr_F}) then one can compute $\hat
G_j= 1/(\hat \chi_j'+ i \hat \chi_j'')$, where $\hat \chi''$ is
computed from Eq.~(\ref{eq:FDR}) with $T_{eff}=T$. Obviously the
real part of $G$ is given by:
\begin{eqnarray}
G_j'(\omega,t_w)= k_j+K_{Lap}(\omega,t_w)={T_{eff}(\omega,t_w) \over
T} \hat G_j'(\omega,t_w).
 \label{eq:G}
\end{eqnarray}
Using the two measurements at $k_1$ and $k_2$ one can solve for
$K_{Lap}$ and $T_{eff}$ getting:
\begin{eqnarray}
K_{Lap}&=& { k_1   \ \hat G'_2(\omega,t_w)- k_2  \  \hat
G'_1(\omega,t_w) \over  \hat G'_1(\omega,t_w)- \hat
G'_2(\omega,t_w)}, \\
T_{eff}(\omega,t_w)&=& T \left({k_1- k_2 \over \hat
G'_1(\omega,t_w)- \hat G'_2(\omega,t_w) }\right).
\end{eqnarray}
 It is clear that if one finds  a dependence of $T_{eff}$ on
$\omega$ this method cannot be used because $\chi''(\omega,t_w)$ is
not simply proportional to $S_j(\omega,t_w) \omega$ as assumed in
Eq.~(\ref{eq:KrKr}). This method is very sensitive to the shape of
the spectrum and to its statistical noise.  Thus, before computing
the elastic modulus, we smooth the spectra to reduce the statistical
uncertainty.

\subsection{Active microrheology}

Active microrheology (AMR) involves the active manipulation of probe
particles by an external force exerted by optical tweezers. In our
case, we apply an oscillatory force $f_0(t, \omega)$ at certain
frequencies $\omega$ by means of the spatial modulation
$x_0(t,\omega)$ of trap '3'. One determines the Fourier transform of
the  linear response function:
\begin{equation}\label{eq:ftchi}
    \hat{\chi}(\omega,t_w) = \frac{\hat{x}(\omega,t_w)}{\hat{f}_0(\omega)},
\end{equation}
in order to resolve the viscoelastic properties of the colloidal suspension.

As mentioned before, there are a priori two times scales in the
glassy system. The viscoelastic properties evolve in a slow aging
time $t_w$ ($\gamma=6\pi r \eta = \gamma(\omega,t_w)$, $K_{Lap} =
K_{Lap}(\omega,t_w)$) while displacements fluctuations evolve in the
fast time scale $t \ll t_w$. The effective hookean force on the
particle due the relative displacement $x(t)-x_0(t,\omega)$ with
respect to the laser beam focus is $-k_3(x(t)-x_0(t,\omega))$ while
the force due the evolving elasticity of the medium is
$-K_{Lap}(\omega,t_w)x(t)$. Hence, the motion of the probe particle
is described by the following Langevin equation
\begin{equation}\label{eq:langvisco}
    \gamma(\omega,t_w) \dot{x} + k_3(x-x_0(t,\omega)) + K_{Lap}(\omega,t_w)x = \sqrt{2k_B
    T_{eff}(\omega,t_w)
    \gamma(\omega,t_w)}
    \xi(t).
\end{equation}

In order to derive a relation between the linear response function
and the frequency-dependent viscoelastic properties of the colloidal
glass, Eq.~(\ref{eq:langvisco}) must be recasted with no stochastic
term, as
\begin{equation}\label{eq:langvisco1}
    \gamma(\omega,t_w) \dot{x} + (k_3 + K_{Lap}(\omega,t_w))x =  f_0(t,\omega),
\end{equation}
where $f_0(t,\omega) = k_3x_0(t,\omega)$ is the active force term.
The Fourier transform of Eq.~(\ref{eq:langvisco1}) is computed over
a time window of 10 s (such that $\gamma$ and $K_{Lap}$ are
constant) and moving-averaged over 50 seconds. It leads to the
following expression for the inverse of the response function at
given frequency and aging time
\begin{equation}\label{eq:invresp}
    \frac{1}{\hat{\chi}(\omega,t_w)}=i\omega\gamma(\omega,t_w)+(k_3+K_{Lap}(\omega,t_w)),
\end{equation}
Therefore, by measuring directly the mechanical response at a
given frequency of the particle motion to the applied external
force, it is possible to resolve either the relative viscosity and
the elasticity of the colloidal glass during aging by means of the
expressions
\begin{equation}\label{eq:visco}
    \frac{\omega \gamma(\omega,t_w)}{k_3} = Im\{\frac{1}{k_3 \hat{\chi}(\omega,t_w)}\},
\end{equation}
\begin{equation}\label{eq:elast}
    \frac{K_{Lap}(\omega,t_w)}{k_3} = Re\{ \frac{1}{k_3 \hat{\chi}(\omega,t_w)}
    \}-1.
\end{equation}

In our case, the position of trap '3' is oscillated in time along
the $x$ direction at three different frequencies $\omega =
\omega_j, j=1,2,3$ simultaneously according to
\begin{equation}\label{eq:sinmotion}
    x_0(t,\omega)=A(\sin(\omega_1 t)+\sin(\omega_2 t)+\sin(\omega_3 t)),
\end{equation}
where $f_1 = \omega_1 / 2 \pi = 0.3$ Hz, $f_2 = \omega_2 / 2 \pi =
0.5$ Hz, $f_3 = \omega_3 / 2 \pi = 1.0$ Hz, and $A = 9.2 \times
10^{-7}$ m. Higher frequency sinusoidal oscillations ($\omega/2\pi
=$ 2.0, 4.0, 8.0 Hz) were also checked in order to compare our
results at low frequencies to higher ones.

AMR allows to determine directly the effective temperature of the
colloidal glass at given frequency $\omega$ and aging time $t_w$ by
means of Eq.~(\ref{eq:Teff}), as suggested by \cite{jabbari1}. First
of all, one needs to synchronize the input forcing signal
$x_0(t,\omega)$ with the response of the trapped bead $x(t)$. The
Fourier transform of the response function is determined by using
Eq.~(\ref{eq:ftchi}) for a probe particle driven by trap '3': in
practice we divide the power spectral density of the output signal
$|\hat{x}(\omega,t_w)|^2$ by the corresponding transfer function
$\hat{x}^*(\omega,t_w) \hat{f}_0(\omega)$. On the other hand, one
needs to determine the power spectral density of the displacement
fluctuations in the absence of external forcing but for the same
value of the trapping stiffness. For this reason, traps '2' and '3'
are created with the same stiffness. We measure the power spectral
density of fluctuations of a particle kept by trap '2'.

\section{Experimental results of the single bead experiment }

\subsection{Laser modulation method}\label{res_laser_mod}

\begin{figure}
\begin{center}
\subfigure[]{ \resizebox*{6cm}{!}{\includegraphics{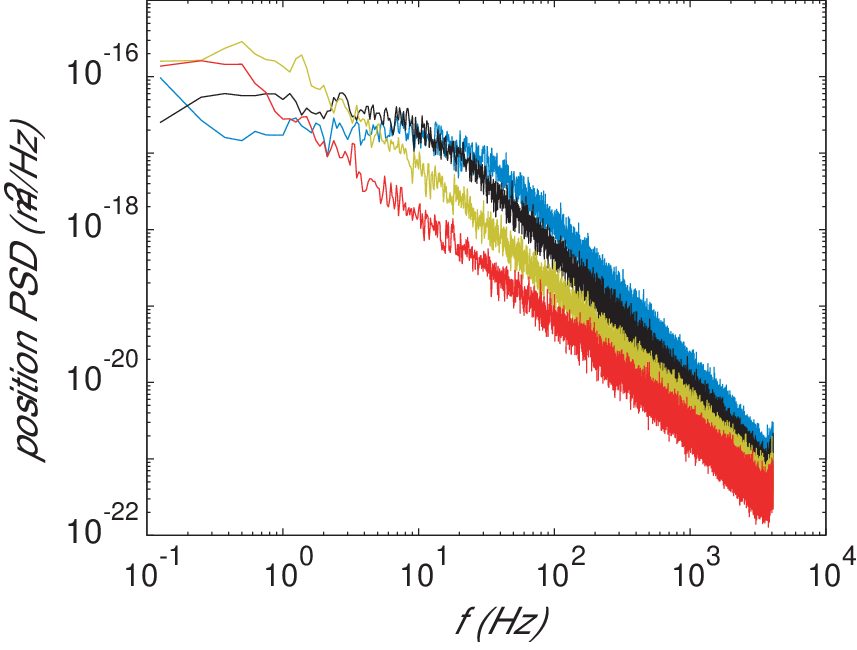}}}
\subfigure[]{
\resizebox*{6cm}{!}{\includegraphics{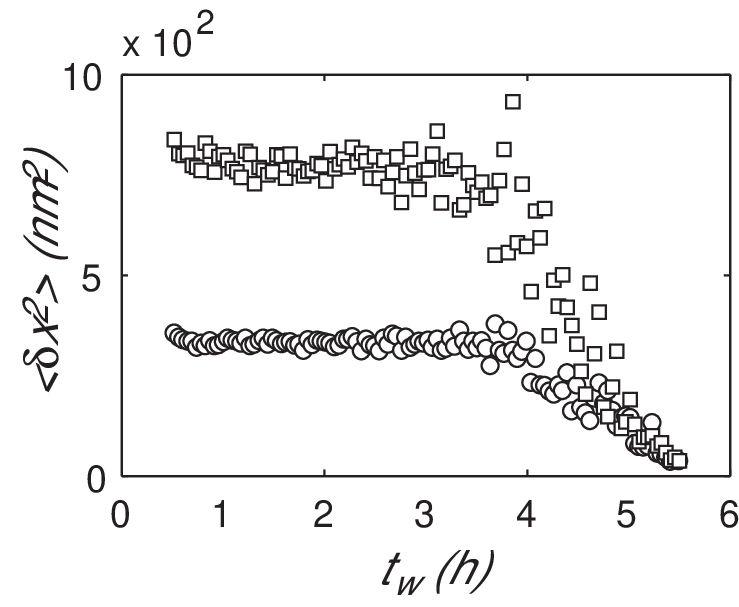}}}
\caption{\label{fig:spectra} (a) Spectra of the particle position
fluctuations inside a Laponite solution at $2.3\%$ wt. The spectra
has been measured for the trap stiffness $k_1=6.34$ pN/$\mu$m
 at various ageing times: from right to left
$t_w=30$,  $110$, $190$ and $270$ minutes. (b) Evolution of
$\left<\delta x^2\right>$ with $t_w$ for both stiffness, $k_1=6.34$
pN/$\mu$m  ($\square$) and $k_2=14.4$ pN/$\mu$m, ($\circ$). The
variances have been measured on a time window of { $\tau= 61$ s.}}
\end{center}
\end{figure}

\begin{figure}
\begin{center}
{\resizebox*{13cm}{!}{\includegraphics{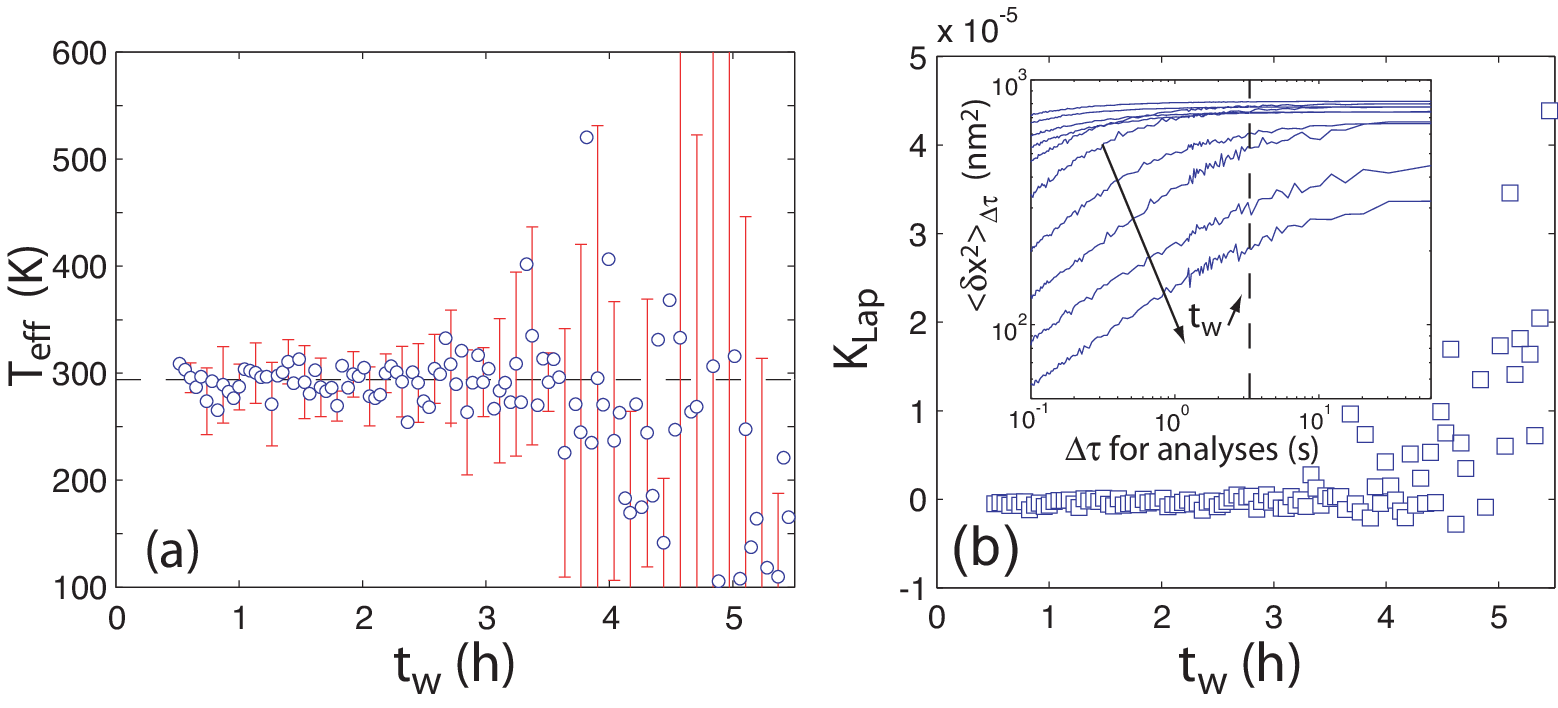}}}
\caption{\label{fig:Teff} (a) Evolution of the effective temperature
with aging time for the data of  Fig. \ref{fig:spectra} $k_1=6.34$
pN/$\mu$m, $k_2=14.4$ pN/$\mu$m, 2.3\%
 wt of Laponite and { $r=$ 1 $\mu$m} glass bead. The error bars are computed from the statistical error of $\left<\delta x_j^2\right>$ at the time $t_w$.
  (b) Evolution of the stiffness of the Laponite with ageing time.
  Inset: Evolution of $\left<\delta x^2\right>_{\Delta \tau}$ for { $k_1 = 6.34$ pN$/\mu$m} as a function of the duration $\Delta \tau$
  of the samples for different aging times : $t_w=30$, $57$, $83$, $110$, $137$, $163$, $190$, $217$, $243$ and $270$ minutes.}
\end{center}
\end{figure}

Fig. \ref{fig:spectra}(a) shows the power spectra of the particle
fluctuations inside Laponite at concentration measured at four
different $t_w$ with the trap stiffness $k_1=6.34$ pN/$\mu$m. We see
that  as time goes on the low frequency component of the spectrum
increases. That is the cut-off frequency $(k_j+K_{Lap})/\gamma$
decreases mainly because of the increasing of the viscosity.  At
very long time this cut-off is well below 0.1 Hz.
In Fig. \ref{fig:spectra}(b) we plot the variance of the particle
measured for the same data of Fig. \ref{fig:spectra}(a) on time
windows of length $\tau=61$ s. The variances remain constant for a
very long time and they begin to decrease because of the increase of
the gel stiffness. Using these data and Eq.~(\ref{eq:temperature})
and Eq.~(\ref{eq:KLap}) one can compute $T_{eff}$ and $K_{Lap}$. The
results for $T_{eff}$ and $K_{Lap}$ are shown in Fig.
\ref{fig:Teff}(a) and  \ref{fig:Teff}(b) respectively.

 We find that $T_{eff}$ is constant at the beginning and is very close to $T_{bath}=294$ K,
 then when the jamming occurs, that is when $K_{Lap}$ increases,
 it becomes more scattered without any clear increase with $t_w$, contrary to Ref.
\cite{greinert}. We now make several remarks. First, we point out
that the uncertainty of their results are underestimated. The error
bars in Fig.~\ref{fig:Teff}(a) are here evaluated from the standard
deviation of the variance using Eq.~(2) in Ref \cite{greinert} at
the time $t_w$. Although they are small for short time $t_w$,
($\Delta T_{eff}/T_{eff}\leq 10\%$), they increase for large $t_w$.
This is a consequence of the increase of variabilities of
$\left<\delta x_j^2\right>$ as the colloidal glass forms. This point
is not discussed in in Ref. \cite{greinert} and we think that the
measurement errors are of the same order or larger than the observed
effect. The results depend on the length of the analyzing time
window and the use of the principle of energy equipartition becomes
questionable for the following reasons.

First, these analyzing windows cannot be made too large because
the viscoelastic properties of Laponite evolve as a function of
time.
 Second, the corner frequency of the global trap (optical trap and gel),
 the ratio of the trap stiffness to viscosity,
 decreases continuously mainly because of the increase of viscosity.
 At the end of the experiment,
 the power spectrum density of the displacement of the bead shows that
 the corner frequency is lower than 0.1 Hz.
We thus observe long lived fluctuations, which could not be taken
into account with short measuring times. This problem is shown on
Fig.~\ref{fig:Teff}(b). We split our data into equal time duration
$\Delta \tau$, compute the variance and average the results of all
samples. The dotted line represents the duration 3.3 s chosen in
\cite{greinert}. At the beginning of the experiment, the variance of
the displacement is constant for any reasonable durations of
measurement. However, we clearly see that this method produces an
underestimate of $\left<\delta x^2\right>$ for long aging times,
specially when the viscoelasticity of the gel becomes important.
Long lived fluctuations are then ignored.

\begin{figure}
\begin{center}
\subfigure[]{
{\resizebox*{6cm}{!}{\includegraphics{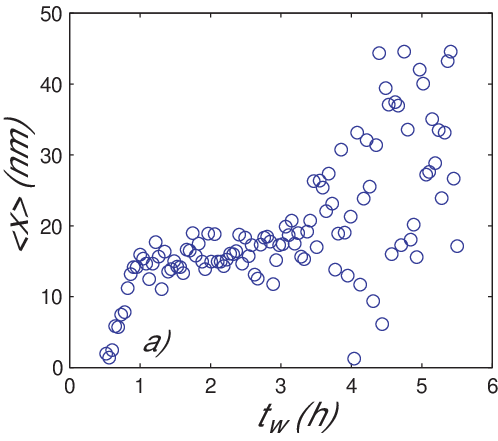}}}}
\subfigure[]{
{\resizebox*{6cm}{!}{\includegraphics{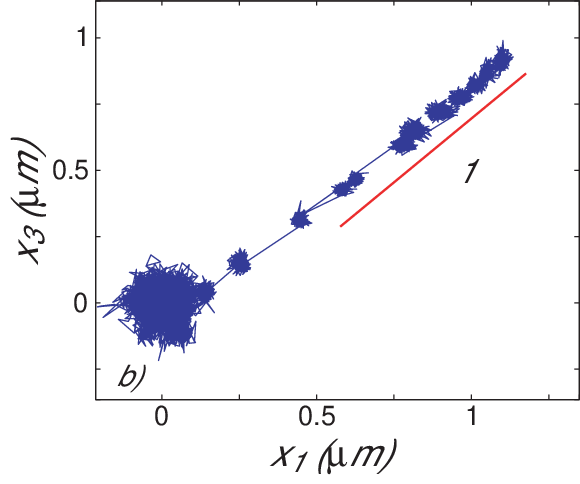}}}}\\
\caption{\label{fig:drift} (a) Evolution of the mean position along
the $x$-direction with aging time, same data as in Fig.
\ref{fig:spectra}. (b) Comparison of the trajectories for { two $1
\mu$m in diameter glass beads of } whose fixed traps are separated
by 6.5 $\mu$m during the evolution of the suspension of Laponite
({$I=5 \times 10^{-3}$ mol l$^{-1}$, 1.2\% wt of Laponite). Each
pincushion represents 1 second of measurement at 288 Hz, repeated
every 5 min during 150 min.}}
\end{center}
\end{figure}

Moreover, in some experiments, we observe a small drift of the probe
position at the end of the measurements, when the gel stiffness
becomes very large compared to the optical trap stiffness. Figure
\ref{fig:drift}(a) shows an example of the time evolution of the
mean position of a single trapped bead. The first drift is
associated to the evolution of the temperature inside the sample due
to the laser. After 3.5 hours, the mean position of the bead is less
accurate, which is the consequence of the increase of the viscosity
of the Laponite as mentioned above, but we can observe a drift
toward larger values of $x$. This phenomenon can be enhanced, using
smaller beads in a Laponite sample at higher concentration and
higher ionic strength. We simultaneously measure the positions of
three beads of 1 $\mu$m in diameter in three fixed traps. The beads
start escaping their trap after 90 min. In Figure
\ref{fig:drift}(b), we compare the trajectory of the first particle
with the third one. Even if the trajectories are not correlated over
short times (1 s for a pincushion), at longer time, the drift is
identical for both beads. This global drift is in this case
obviously an artifact that could be due to a small leak or a small
bubble. We point out here that data needs careful analysis to get
rid of strange behavior.

\subsection{Kramers-Kroening and modulation technique} \label{section_Kramers}
\begin{figure}
\begin{center}
\resizebox*{6cm}{!}{\includegraphics{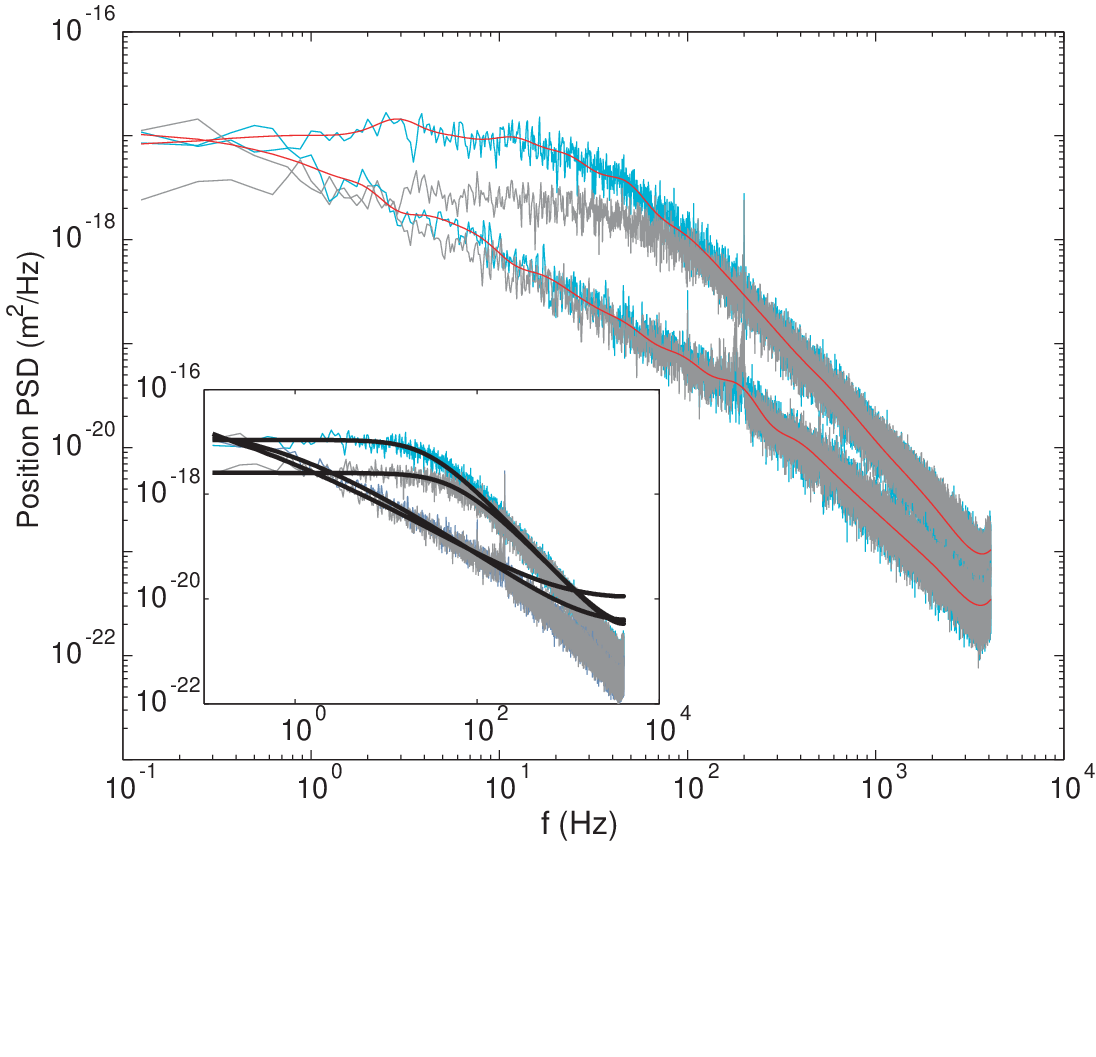}}
\caption{\label{fig:smoothspec} {Example of spectral data and smooth
spectra used with the Kramers-Kroenig relation ($k_1$ = 7.47
pN/$\mu$m, ${k_2}$ = 16.7 pN/$\mu$m  and for two ageing times:
${t_w=}$ 30 min and 240 min) The red smooth lines of the main plot
are obtained by averaging the spectra over fixed windows. Inset: We
check that the position PSD deduced after data treatment (black
thick lines) agree with the noisy spectra up to $10^2$ Hz (see
text).}}
\end{center}
\end{figure}

 This new method, using Kramers-Kronig
relation, allows us to test the dependence of the effective
temperature on the frequency. As we already mentioned,  before
applying Eq.~(\ref{eq:KrKr_F}) to the spectral data we proceed to a
smoothing of the spectra $S_j(\omega)$. This operation is shown in
Fig. \ref{fig:smoothspec}. Then, instead of directly computing $\hat
\chi_j'(\omega,t_w)$ using Eq.~(\ref{eq:KrKr_F}), we use the
derivative of the inverse Fourier transform of the smoothed spectra.
The artificial undulations of the smoothed spectra produce
fluctuations on the final curves. To decrease this effect, we
additionally fit the inverse Fourier transform by a stretched
exponential function. These methods give finally results that are
less sensitive to the noise of the spectra. We also check that this
method produces spectra in agreement with the first ones (inset of
Fig. \ref{fig:smoothspec}) up to the limit of 100 Hz.

\begin{figure}
\begin{center}
\subfigure[]{
\resizebox*{6cm}{!}{\includegraphics{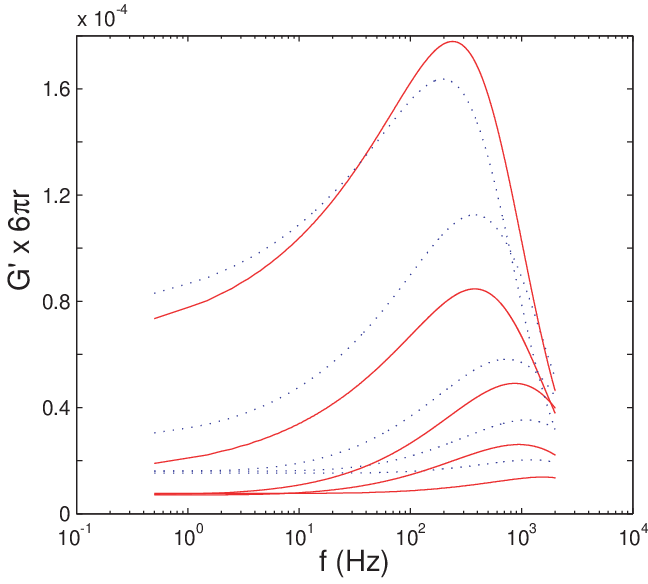}}}
\subfigure[]{ \resizebox*{6cm}{!}{\includegraphics{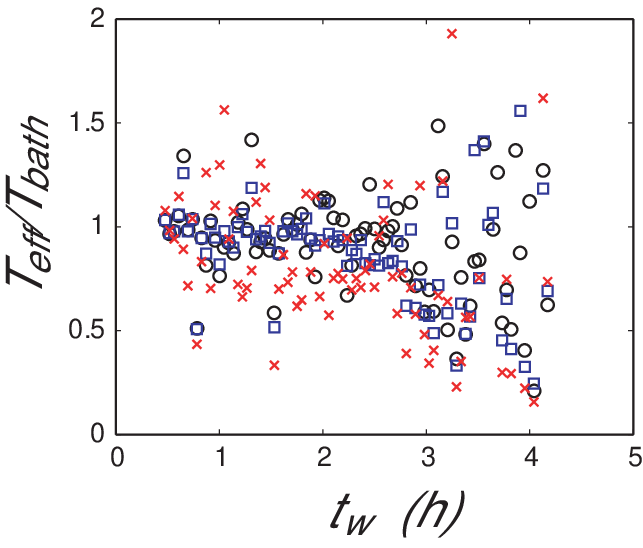}}}
\caption{\label{fig:krkr}(a) {Evolution of $6\pi rG'$ (the global
elastic modulus)} as a function of frequency for both trap
stiffnesses, $k_1=7.47$ pN/$\mu$m (solid lines), $k_2=16.7$
pN/$\mu$m (dashed lines) and for increasing { aging} time: $t_w=30$
min, $90$ min, $140$min, $190$ min and $240$ min (1.2\% wt of
Laponite, 2 $\mu$m glass bead and ionic strength { $I=5 \times
10^{-3}$ mol l$^{-1}$}).
  (b) Evolution of the ration $T_{eff}/T_{bath}$ with aging time ($t_w$)
for different frequencies ({$f=1$} Hz $\square$, 10 Hz $\circ$ and 100
Hz $\times$)).}
\end{center}
\end{figure}

Figure \ref{fig:krkr}(a) shows the real part of $G$, which
corresponds to the global elastic modulus of the gel and of the
laser, for both trap stiffness. The increase with time of $G'$ is
consistent with the increase of the strength of the gel. The elastic
behavior of Laponite is also more pronounced at high frequency. The
last decrease of the curves at very high frequencies is due to the
numerical method. Indeed, the frequency cutoff should be set at
least a  decade below the frequency of the data acquisition: data
above 200 Hz is not reliable. We see that the curves of each
stiffness are well separated except at the end of the measurement,
where the results are not accurate due to the large difference
between the optical stiffness and Laponite stiffness.

From these data, we compute the ratio of the effective temperature
to the bath temperature along the ageing process. These results are
shown on Fig. \ref{fig:krkr}(b) for three different frequencies
($f=1$ Hz, 10 Hz and 100 Hz). We first note that the three curves
are almost identical. This means that the effective temperature does
not depend on frequency. Second, the temperature ratio is close to
1. The dispersion of the data is rather small at early times and
again increases when the stiffness of the gel overcomes those of the
optical traps. This comes from the uncertainties of the elastic
modulus which become larger than the difference between the two
curves. This dispersion may also give negative temperatures for very
long aging time, not shown in Fig. \ref{fig:krkr}(b) which is an
expanded view. Even if this method is
 less accurate than the previous one, it allows us to verify
that the effective temperature is the same for all frequencies.

\section{Results of the multiple trap experiment}

In this section, unless otherwise stated, we will describe only the
results obtained at 2.8 \% wt concentration of Laponite. Indeed at
this concentration the Laponite gelation occurs in a few hours.

\subsection{Viscoelastic properties of the aging colloidal glass}

\begin{figure}
     \centering
     \subfigure[]{
          \includegraphics[width=.31\textwidth]{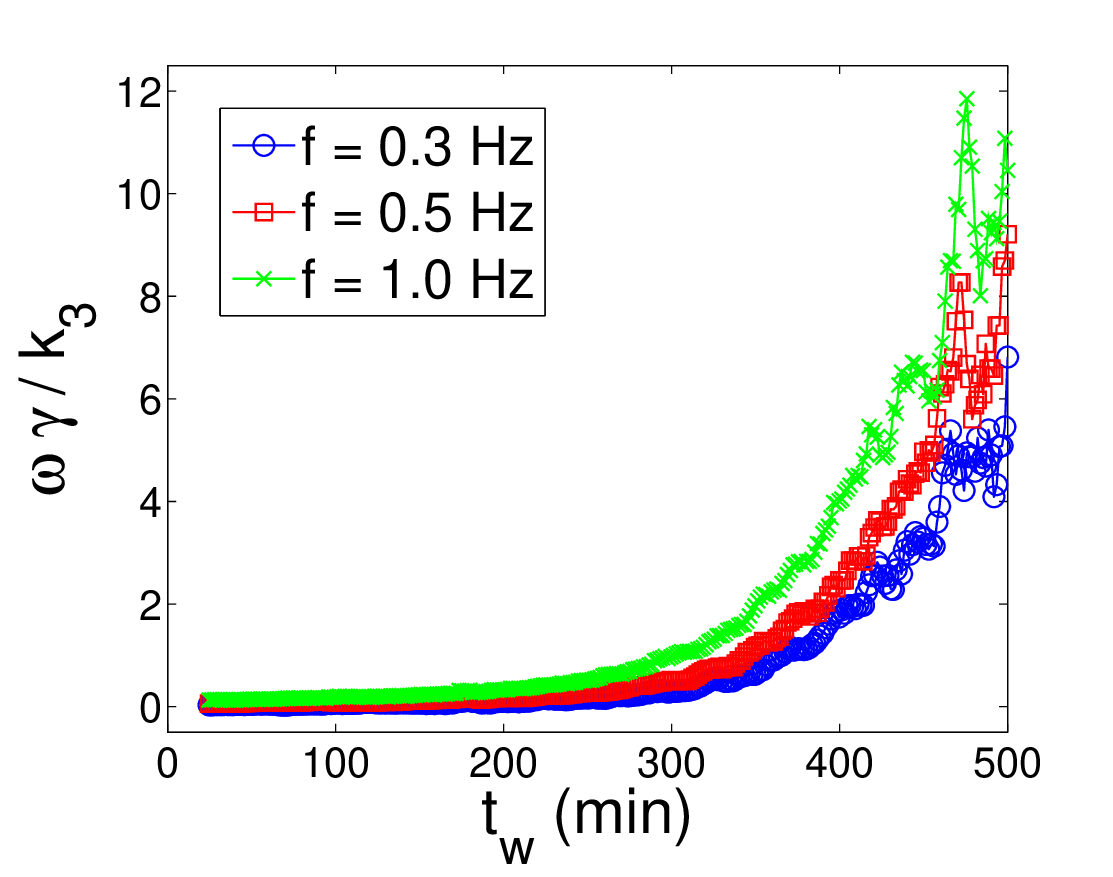}}
     \hspace{.0in}
     \subfigure[]{
          \includegraphics[width=.31\textwidth]{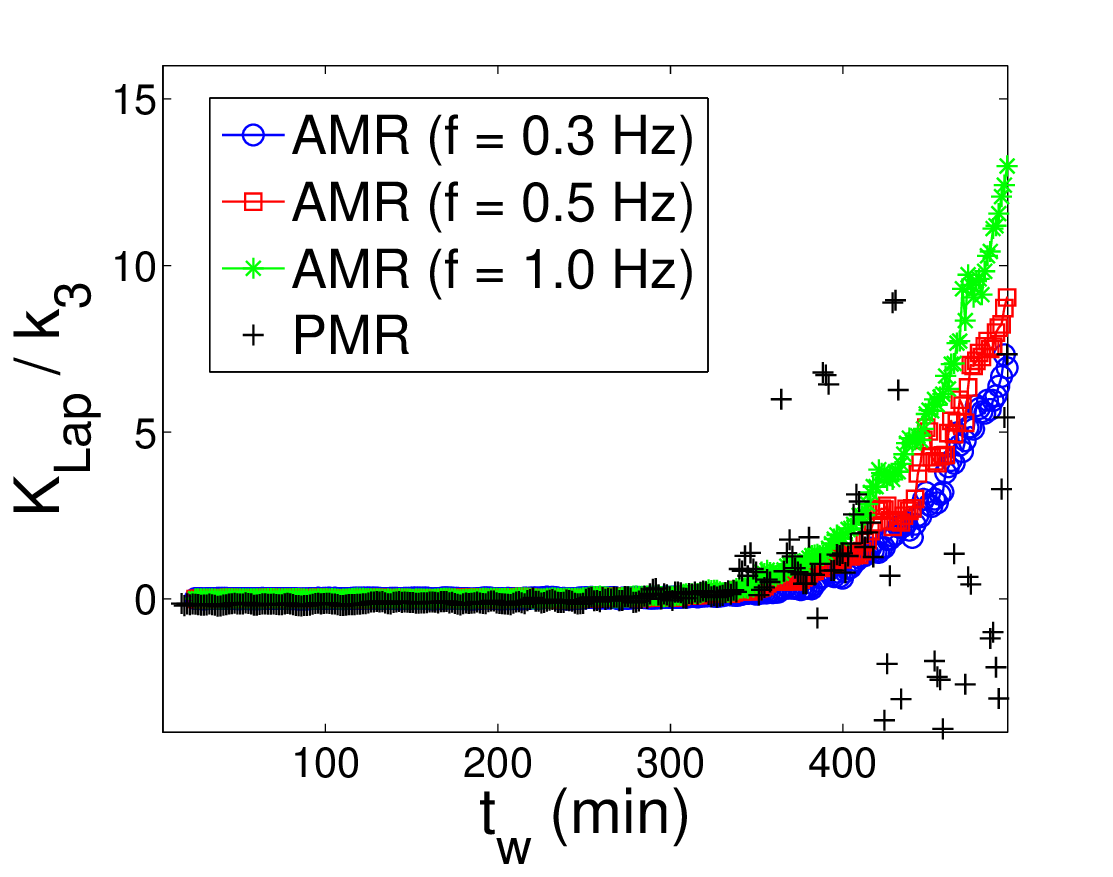}}
     \hspace{.0in}
     \subfigure[]{
          \includegraphics[width=.31\textwidth]{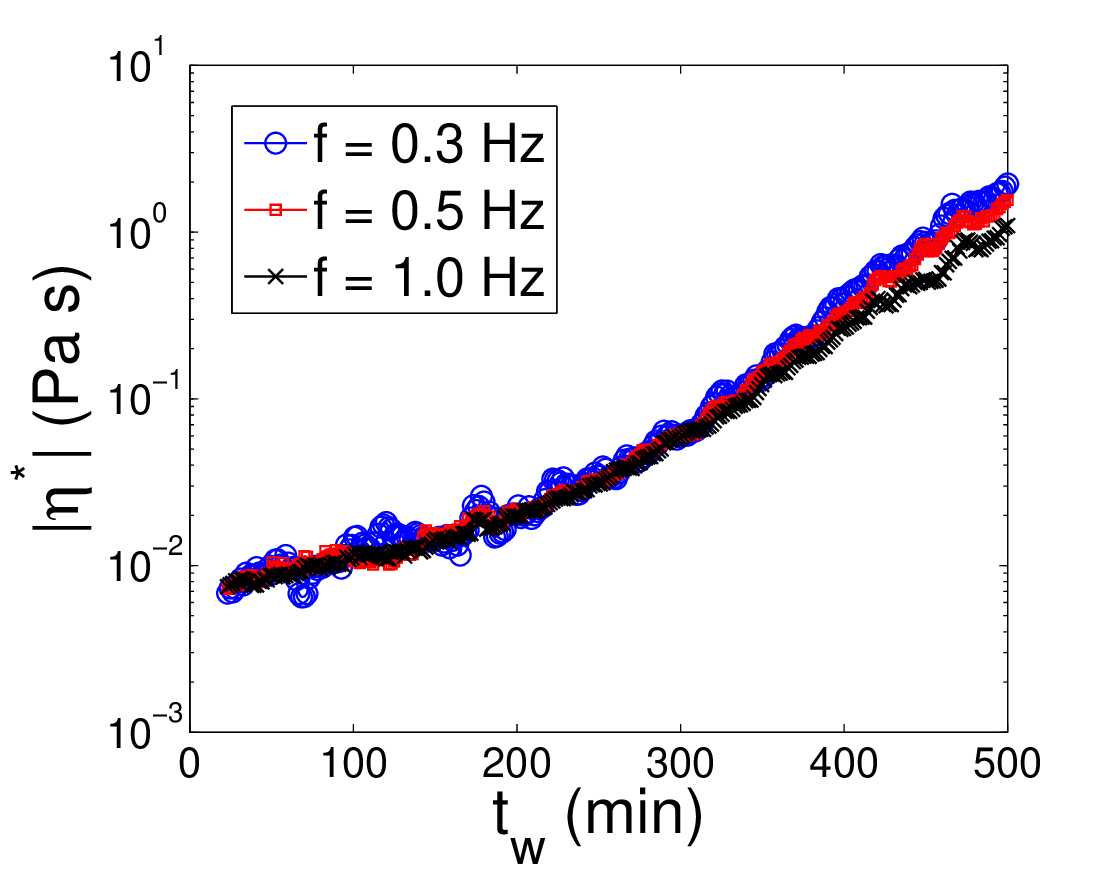}}\\
     \caption{ (a) Time evolution of viscosity of the colloidal glass obtained
     by means of AMR. (b) Time evolution of elasticity of the colloidal glass
     obtained either by PMR and AMR.
     (c) Time evolution of the modulus of the complex viscosity}
     \label{multifig2}
\end{figure}

We first present separately the results of the time evolution of
viscosity and elasticity of the colloidal glass during aging. The
time evolution of the dimensionless quantity $\omega \gamma / k_3$,
linearly proportional to the dynamic viscosity $\eta$ of the
colloidal glass, is shown in Fig. \ref{multifig2}(a). As expected,
it increases continuously as the system ages. On the other hand, the
evolution of the stiffness $K_{Lap}$ is qualitatively different, as
shown in Fig. \ref{multifig2}(b). For $t_w < 300 $ min, $K_{Lap}
\approx 0$, revealing an entirely viscous nature of the laponite
suspension, while for $t_w> 300$ min it becomes viscoelastic, with
$K_{Lap}$ increasing dramatically in aging time from 0 to 10 times
the stiffness of the second trap $k_2 = 7.15$ pN/$\mu$m. Both PMR
and AMR lead to consistent and complementary results and we notice
that AMR measurements are more accurate, leading to a very small
dispersion of data around the mean trend. Instead, PMR measurements
become very sensitive to the inverse of the difference $\langle
\delta x^2 \rangle_1 - \langle \delta x^2 \rangle_2$ as $t_w$
increases, leading to increasingly large data dispersion for $t_w >
350$ min. In order to compare our AMR results with previous
rheological measurements, in Fig. \ref{multifig2}(c) we plot the
time evolution of the modulus of the complex viscosity, given by
$|\eta^*| = (\gamma^2 + K_{Lap}^2/\omega^2)^{1/2}/6\pi R$. We
observe that $|\eta^*|$ increases, almost exponentially,  of  two
orders of magnitude during the first 500 minutes of aging. The
behavior of $|\eta^*|$ is in good agreement with previous
rheological measurements \cite{bonn1}.

\subsection{Properties of non-equilibrium fluctuations during
aging} \label{section_properties} 
\begin{figure}
     \centering
     \subfigure[]{
          \includegraphics[width=.44\textwidth]{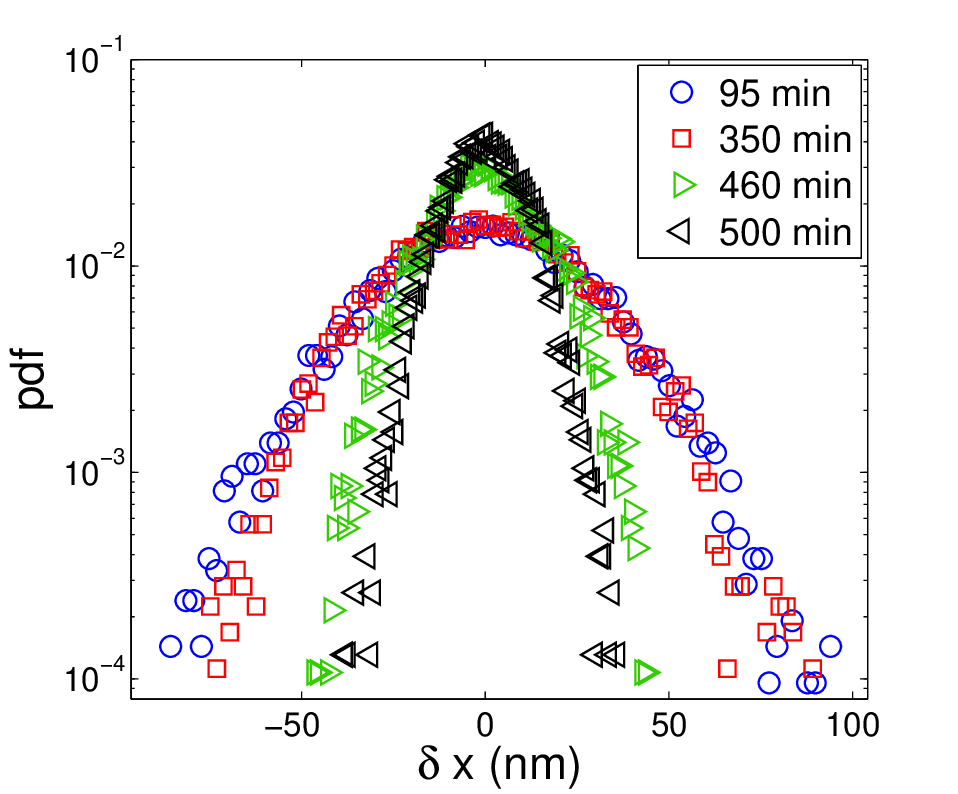}}
     \hspace{.0in}
     \subfigure[]{
          \includegraphics[width=.44\textwidth]{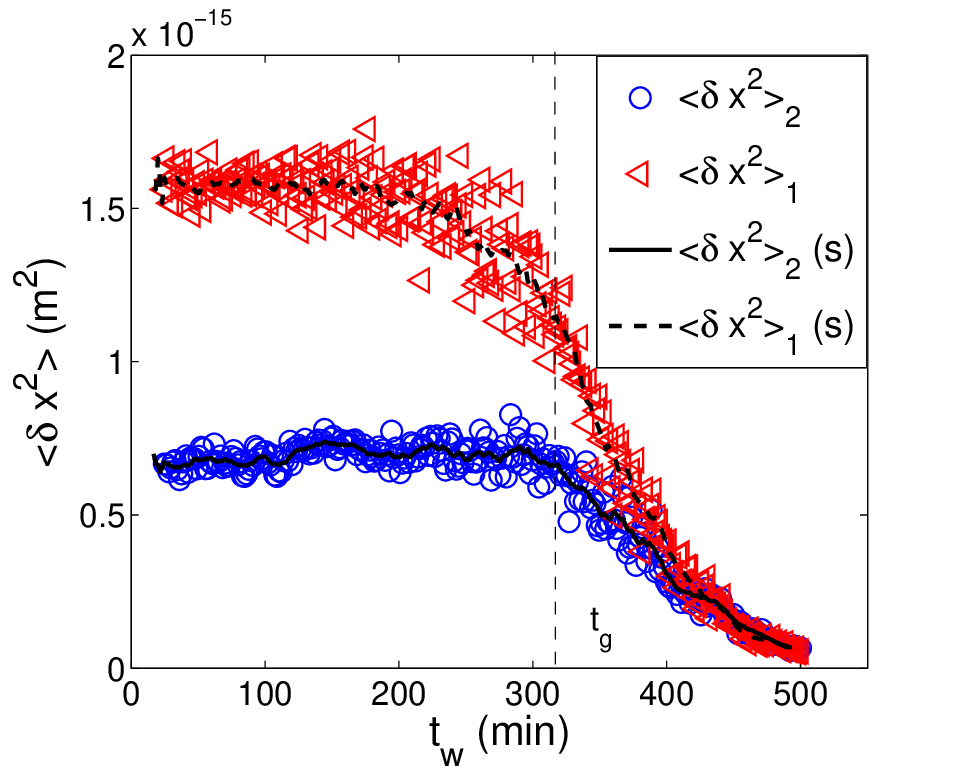}}\\
     \vspace{.0in}
     \subfigure[]{
          \includegraphics[width=.44\textwidth]{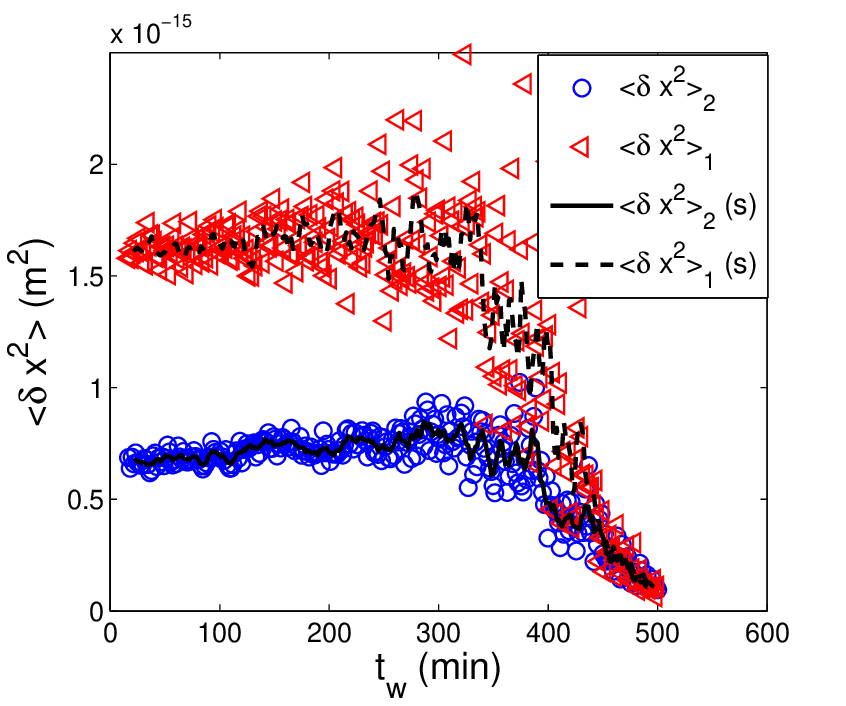}}
     \hspace{.0in}
     \subfigure[]{
          \includegraphics[width=.44\textwidth]{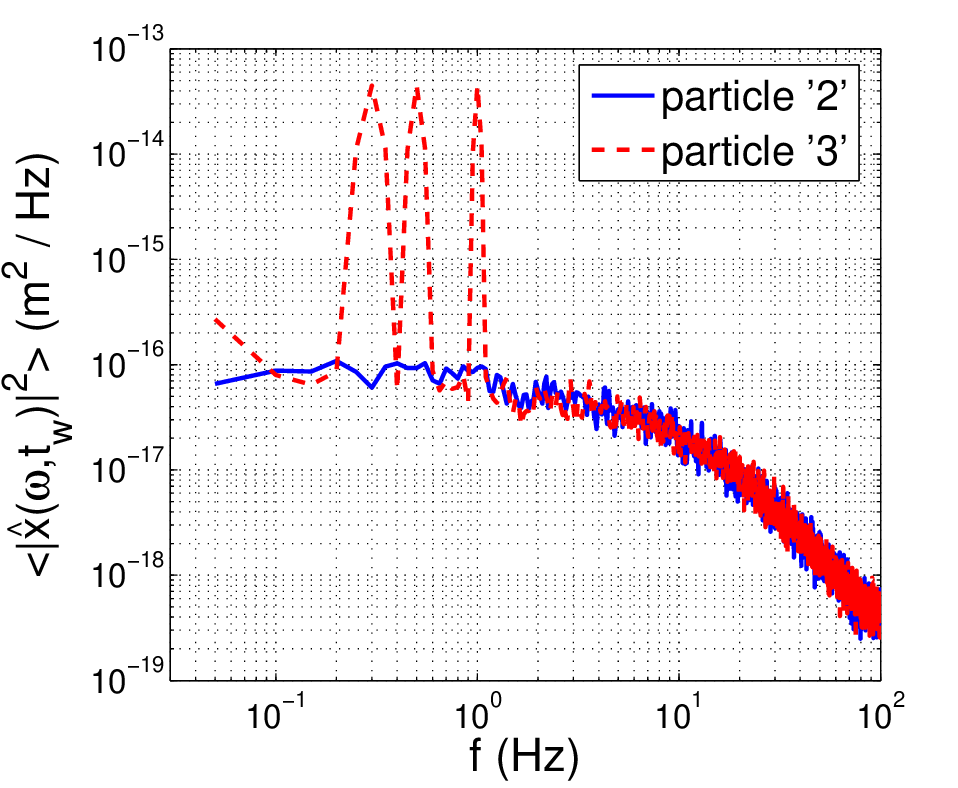}}\\
     \vspace{.0in}
     \subfigure[]{
          \includegraphics[width=.44\textwidth]{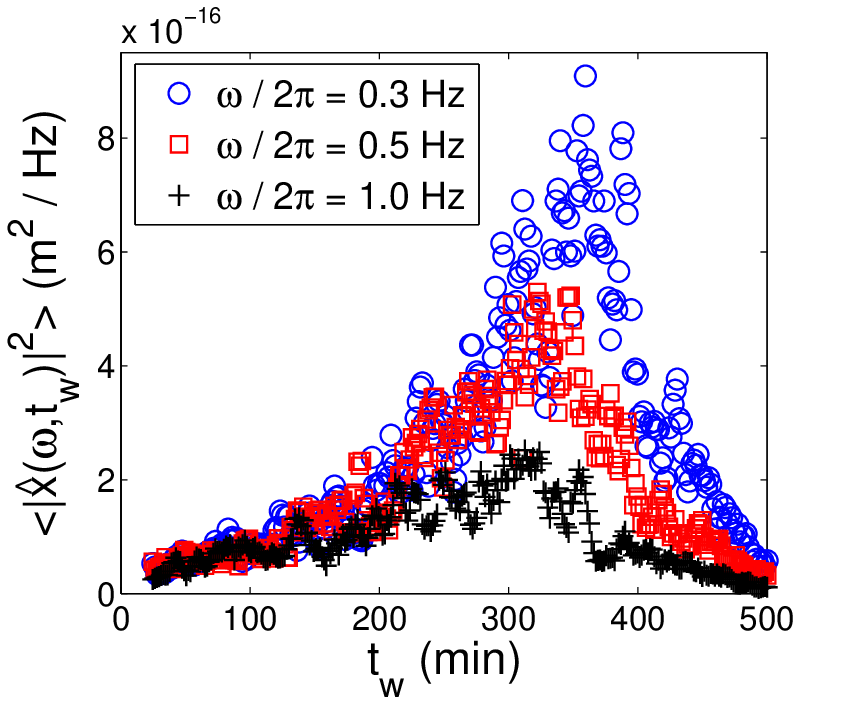}}
     \hspace{.0in}
     \subfigure[]{
          \includegraphics[width=.44\textwidth]{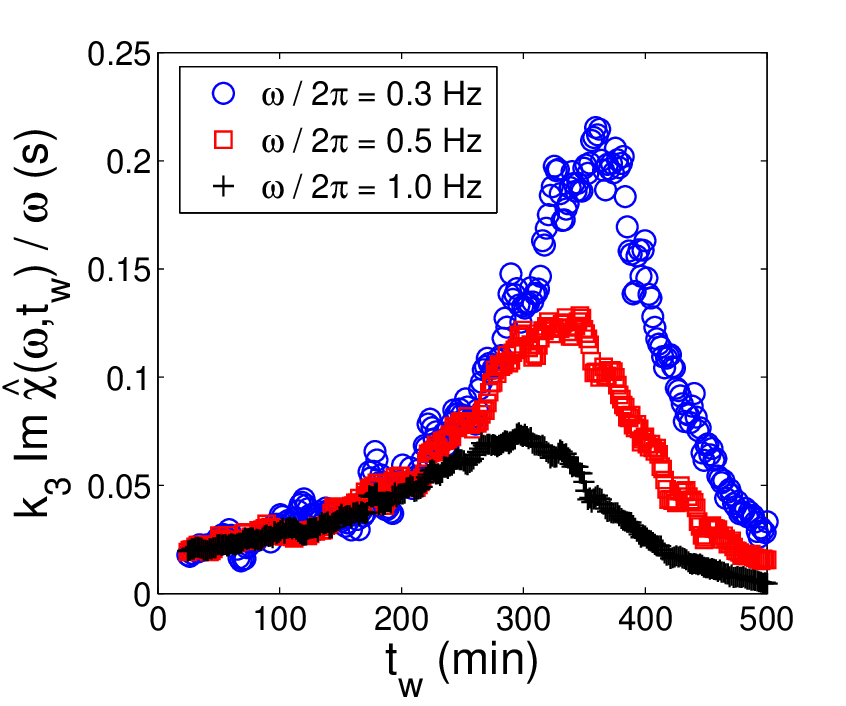}}\\
     \caption{ (a) PDFs of $\delta x$ at
     different $t_w$ for particle '2'. (b) Time evolution
     of variances for particles '1' and '2' computed for $\Delta \tau= 3$ s. The thin dashed line
     indicates the approximate aging time when gelation begins to take place. Labels (s)
     correspond to smooth curves (see text). (c)
     Time evolution of variances computed for $\Delta \tau= 25$ s.
     (d) Spectra of displacement fluctuations of particles '2' and '3'
     at $t_w =$ 95 min.
     (e) Time evolution of the power spectral densities of displacement
     fluctuations of particle '2' for the three frequencies studied.
     (f) Time evolution of the imaginary part of the Fourier transform of the
     response functions of particle '3'.}
     \label{multifig3}
\end{figure}

The probability density functions
of displacement fluctuations around their mean positions $\delta x =
x - \langle x \rangle$, computed for particles '1' and '2' over an
analyzing time window $\Delta \tau =$ 25 s, are Gaussian, as shown
in Fig. \ref{multifig3}(a) for particle '2'. The corresponding
variances $\langle \delta x^2 \rangle_1$ and $\langle \delta x^2
\rangle_2$ decrease during aging due to the combined effects of the
increasing viscosity and elasticity of the colloid constraining the
motion of the probe particles. The values of variances depend on the
length of the analyzing time window $\Delta \tau$ and for very short
windows, they become largely underestimated as the system evolves,
as discussed in { Subsection} \ref{res_laser_mod}.

In order to go deeper into the artifacts that can arise for PMR from
the use of very short windows, we analyzed the experimental data
using for two different values $\Delta \tau=$ 3 s and 25 s. Fig.
\ref{multifig3}(b) shows the results for $\Delta \tau=$ 3 s. Two
regimes can be identified like in previous PMR experiments
\cite{greinert}, \cite{jop}: for $t_w \lesssim 300$ min, both
variances are quite constant, while for $t_w \gtrsim 300$ min they
decrease up to one order of magnitude at $t_w = 500$ min. The
transition between one regime to the other is kind of abrupt and
occurs at $t_w \approx 300$ min $\equiv t_g$. This time corresponds
to the transition from a purely viscous liquid-like phase to the
formation of a viscoelastic glassy phase associated to the house of
cards structure, as shown previously. It is noticeable that the
transition point from the plateau to the decaying curve depends
slightly on the value of the stiffness of the optical trap. We
observe that the transition occurs first (around $t_w = 250$ min)
for the weakest trap ('1') than for the strongest one ('2') (around
$t_w = 300$ min), indicating that the motion of a probe particle
could be sensitive to the relative strength of the optical trap with
respect to the elasticity of the colloidal glass close to the
gelation point. However, this is nothing but an artifact due to the
use of a very short analyzing time window. Since the corner
frequency of the power spectral density of displacement fluctuations
depends linearly on the stiffness of the optical trap $k$, large
underestimates of the corresponding variance occur first for the
weakest trap when using $\Delta \tau=$ 3 s, leading to an apparent
dependence of the onset of gelation on $k$. In fact, when using
$\Delta \tau =$ 25 s, there is no such dependence as shown in Fig.
\ref{multifig3}(c). Even when smaller data variability is observed
for $\Delta \tau=$ 3 s than for 25 s, one must be aware that the
latter case is more reliable than the former since longer lived
modes are taken into account in a method which must integrate the
contributions of as many frequencies as possible.

Another artifact can arise easily due to wrong data smoothing.
Figs.~\ref{multifig3}(b) and \ref{multifig3}(c) also show the smooth
curves resulting from the convolution of the data points of the
variances with rectangular time windows of different lengths: 16 min
in Fig.~\ref{multifig3}(b) and 5 min in Fig.~\ref{multifig3}(c).
Wrong estimates become important at late aging times because of the
use of an inappropriate moving window, like in the case of the
window of 16 min.

Both artifacts can lead to an apparent systematic increase of the
effective temperature of the colloid, as shown in Subsection
\ref{subsec:effect_temp}. Hence, one must be very careful when
analyzing experimental data using PMR for a non-ergodic system such
as the colloidal glass studied here.

Frequencies as low as 0.1 Hz could be resolved properly using the
multitrap method, as shown in the spectral curves of displacement
fluctuations of particles '2' and '3' (Fig. \ref{multifig3}(d)). The
application of the oscillating force $f­_0(t,\omega)$ on particle
'3' leads to an increase of two orders of magnitude of its spectrum
at the corresponding frequency $\omega$. Hence, both the power
spectral density and the response function could be measured with
good frequency resolution in the case of AMR. Fig.
\ref{multifig3}(e) shows the time evolution of the power spectral
density for each frequency studied $\omega = \omega_j, j=1,2,3$. The
time behavior of $|\hat{x}(\omega,t_w)|^2$ is completely different
from that of ergodic liquids at equilibrium, whose value is constant
in time. We observe that the nontrivial shape of the power spectral
density has a maximum which depends on the value of the
corresponding frequency. Fig. \ref{multifig3}(f) shows the time
evolution of the imaginary part of the Fourier transform of the
response function at each frequency $\omega = \omega_j, j=1,2,3$,
calculated by means of Eq. (\ref{eq:ftchi}). We observe the same
behavior in time for each frequency, indicating that
$|\hat{x}(\omega,t_w)|^2$ and $Im\{\hat{\chi}(\omega,t_w)\}$ are
related by a proportionality constant during aging, satisfying a
generalized FDT relation (Eq. (\ref{eq:Teff}) with no dependence on
$t_w$), as explained in in the next section. It has to be noted that
the position of these maxima depend on the strength of the optical
traps used to measure the response and fluctuations. These strengths
are the same in our case whereas they are different in the
experiment of Ref. \cite{Abou03} where the fluctuations have been
measured on free particles. This difference in the strength induces
a shift on the time position of these maxima which must be corrected
in the data analysis. Small errors in this correction may of course
induce an anomalous maximum of  $T_{eff}$ as it has been reported in
\cite{Abou03}.

\subsection{Effective temperatures}\label{subsec:effect_temp}

\begin{figure}
     \centering
     \subfigure[]{
          \includegraphics[width=.47\textwidth]{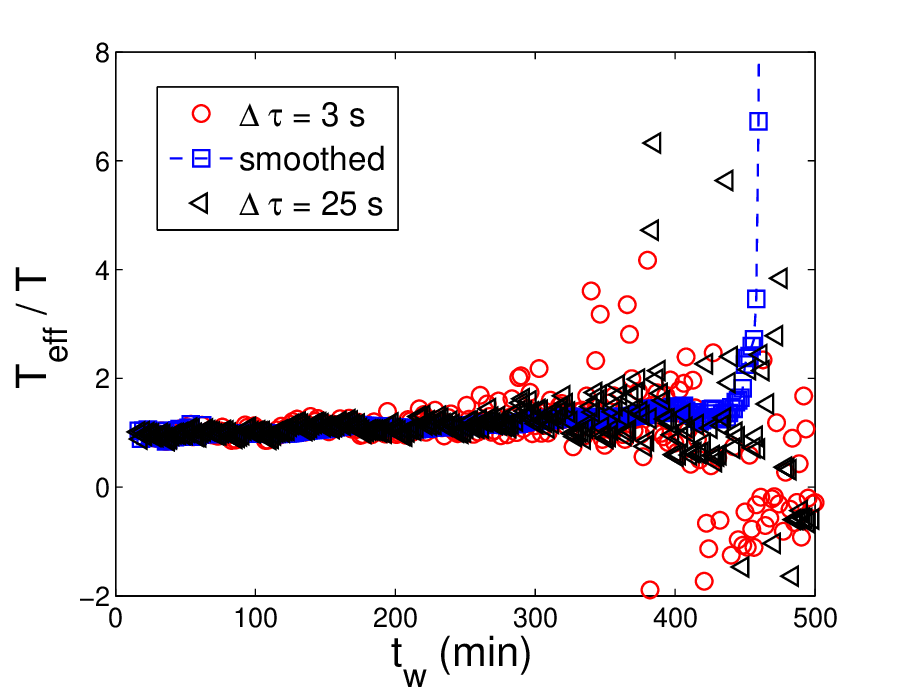}}
     \hspace{.1in}
     \subfigure[]{
          \includegraphics[width=.47\textwidth]{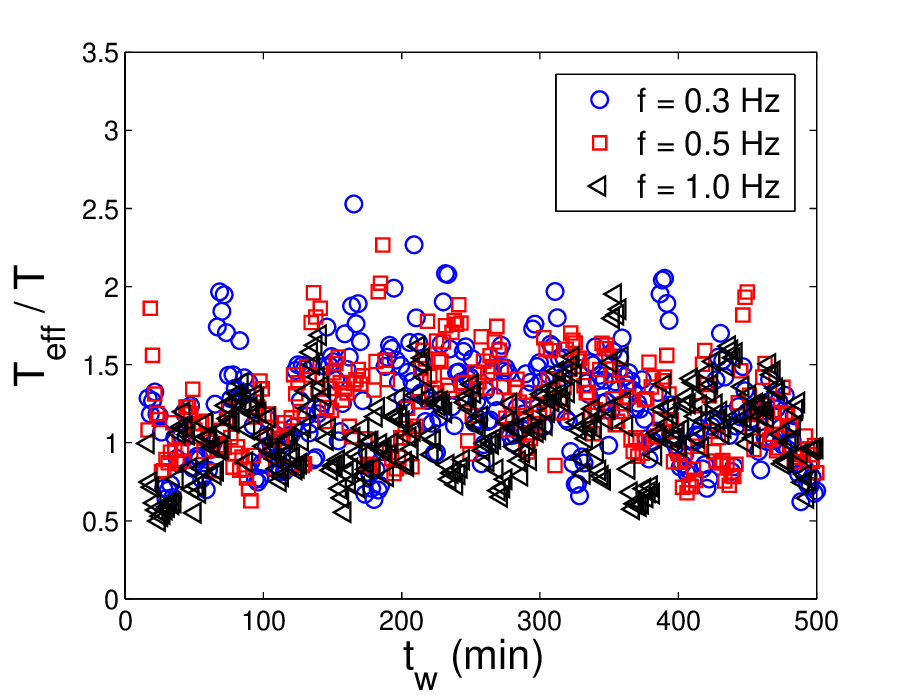}}\\
     \vspace{.1in}
     \subfigure[]{
          \includegraphics[width=.47\textwidth]{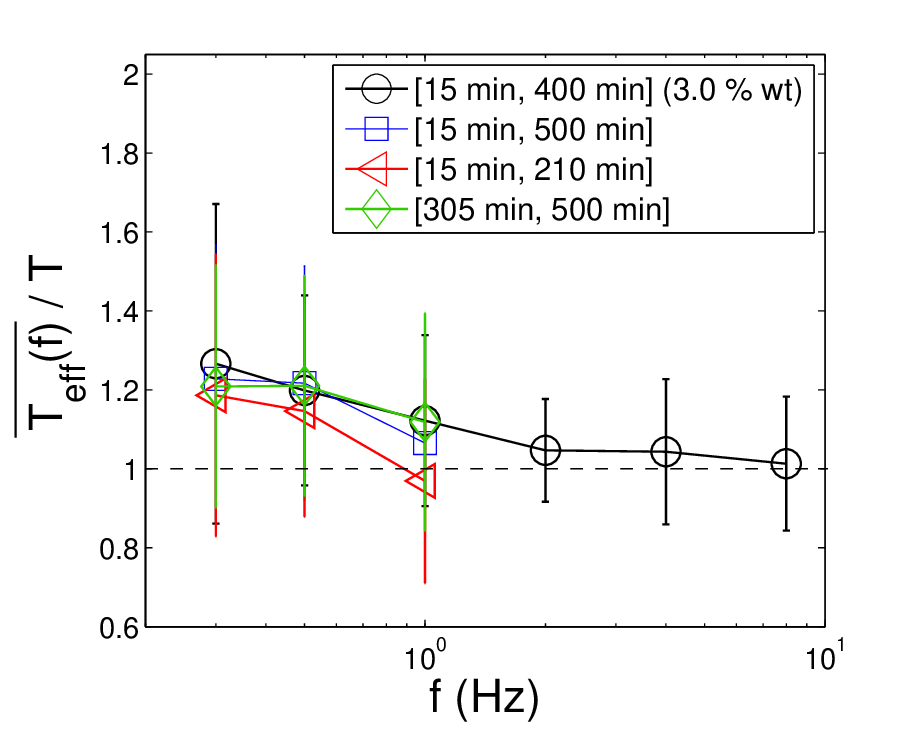}}\\
     \caption{ (a) Time evolution of effective temperatures of the colloid
     during aging obtained by means of PMR (see text for symbols). (b) Time evolution of
     effective temperatures for different frequencies obtained by means of AMR. (c) Aging time
     average of the effective temperatures obtained by AMR for different frequencies.
     Error bars represent the standard deviation of data shown in Fig. \ref{multifig5}(b).}
     \label{multifig5}
\end{figure}

Effective temperatures obtained by means of both PMR and AMR are
shown in Figs. \ref{multifig5}(a) and \ref{multifig5}(b),
respectively. In the case of PMR, for $t_w < 250 $ min the effective
temperature is very close to the bath temperature $T = 295$ K with
very few data dispersion around it. This is due to the fact that at
this aging stage the aqueous laponite suspension behaves as a
viscous liquid leading to small and constant data dispersion of
$\langle \delta x^2 \rangle$. However, for 250 min $<t_w$ the
behavior of $T_{eff}$ depends on the lengths of the moving time
windows used to compute the displacement variances and to smooth
data. As discussed previously, the most reliable results are
obtained by using a sufficiently long analyzing window ($\Delta \tau
=$ 25 s) and a sufficiently short smoothing window (5 min). In this
case no systematic increase of the effective temperature around the
bath temperature is observed even when the suspension has already
gelified (300 min $< t_w <$ 350 min), but data dispersion becomes
very large as $t_w$ increases (triangles in Fig.
\ref{multifig5}(a)). On the other hand, for $\Delta \tau =$ 3 s and
a smoothing window of 5 min, we observe an apparent increase of the
effective temperature for 200 min $< t_w <$ 350 min (circles),
corresponding to the stage when underestimates of the variance occur
first for the weakest trap. Once again, we observe an increasing
variability of the effective temperature leading even to negative
values of $T_{eff}$ due to the fact that the dispersions of $\langle
\delta x^2 \rangle_1$ and $\langle \delta x^2 \rangle_2$ are
comparable to $\langle \delta x^2 \rangle_1 - \langle \delta x^2
\rangle_2$. Finally, for $\Delta \tau =$ 3 s and a smoothing window
of 16 min, a dramatic increase of the effective temperature is
observed for $t_w
>$ 400 min (squares). This is very similar to the behavior of
$T_{eff}$ reported by \cite{greinert} and we assert that such
increase is not an actual violation of FDT but the result of both
the underestimates of the variances and inappropriate data
smoothing.

The AMR results for the effective temperature at different
frequencies are shown in Fig. \ref{multifig5}(b). We verify that
there is no actual systematic increase of $T_{eff}$ as the colloidal
glass ages. The effective temperatures recorded by the probe
particles subject to the non-equilibrium fluctuations in the
colloidal glass at a given time scale $1/\omega$ are equal to the
bath temperature during aging even for the slowest modes studied
($\sim 1/f_1 \approx 3$ s). Unlike PMR measurements, we find that
the variability of the data is constant in aging time, which implies
that AMR is a more reliable method even during gelation. In order to
check if $T_{eff}(\omega,t_w)=T$ within experimental accuracy, by
taking into account the constant behavior of $T_{eff}$ and of the
dispersion around its mean value we can compute the aging time
average of $T_{eff}(\omega,t_w)$ over different aging time intervals
$t_w \in [t_w^i,t_w^f]$
\begin{equation}\label{eq:Teffave}
    \overline{T_{eff}(\omega)}
     = \frac{1}{t_w^f - t_w^i}\int_{t_w^i}^{t_w^f} T_{eff}(\omega,t_w')dt_w',
\end{equation}
for every frequency $\omega$. Fig. \ref{multifig5}(c) shows the
results of $\overline{T_{eff}(\omega)}$ with their respective error
bars corresponding to the standard deviations of the data sets. For
comparision, we also determined the effective temperature of the
colloidal glass at 3.0\% wt of Laponite at higher frequencies ($f=$
2.0, 4.0 and 8.0 Hz) without observing any deviation of the
effective temperatures from the bath temperature within experimental
accuracy (Fig. \ref{multifig5}(c)). We conclude that there is no
violation of FDT for this aging non-equilibrium system with its
effective temperatures equal to the bath temperature, regardless of
the measured time scale.

\subsection{Probability density functions of heat fluctuations}
\label{section_PDF_heat}

\begin{figure}
     \centering
     \subfigure[]{
          \includegraphics[width=.47\textwidth]{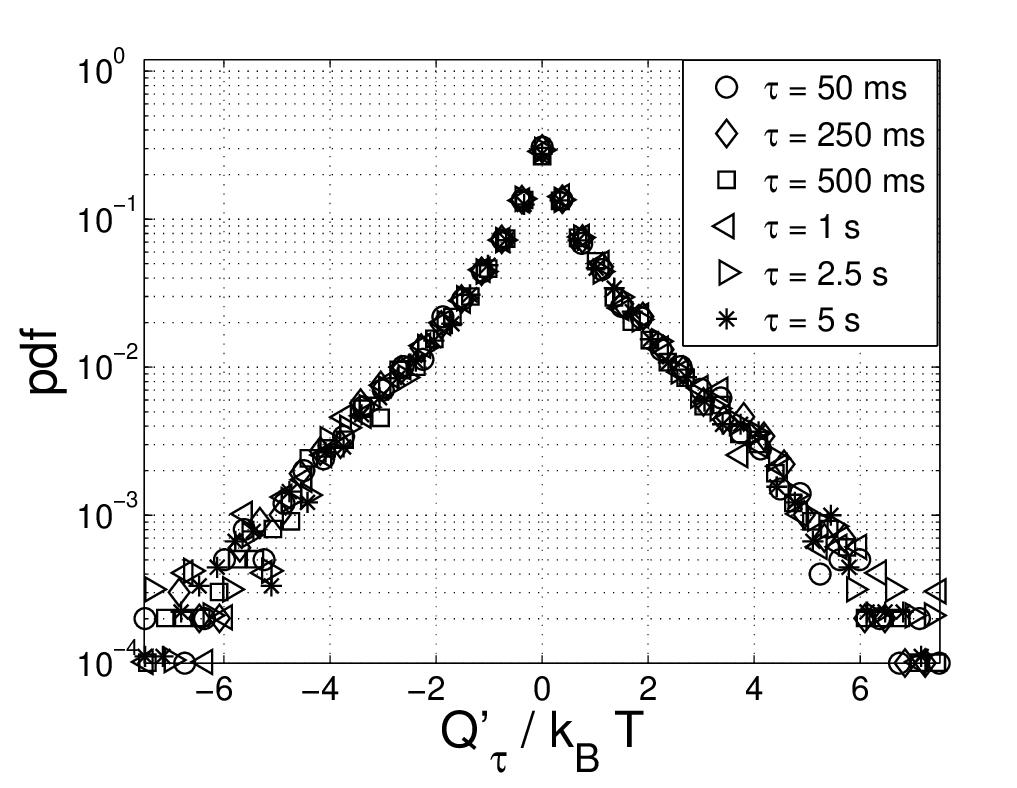}}
     \hspace{.1in}
     \subfigure[]{
          \includegraphics[width=.47\textwidth]{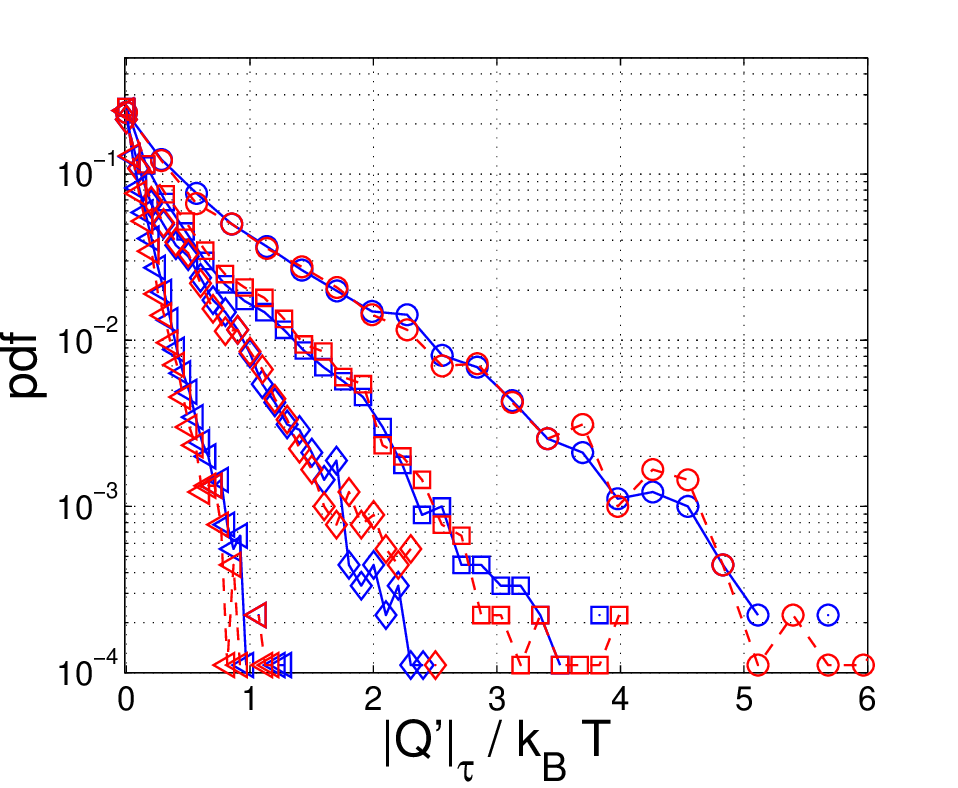}}\\
     \caption{ (a) PDFs of $Q'_{\tau}$ at $t_w = 95$ min for
     different values of $\tau$. (b) PDFs of $|Q'_{\tau}|$ for
     $\tau = 5$ s at different aging times $t_w$ after the onset of gelation {
     (from top to bottom: $t_w = 340$, $435$, $470$ and $500$ min).}
     Blue points correspond to $Q'_{\tau} > 0$ while red ones to $Q'_{\tau} < 0$.}
     \label{multifig6}
\end{figure}

Finally, we present the results for the probability density
functions (PDF) of heat transfers between the colloidal glass and
the surroundings. This kind of analysis has been motivated by the
recent experimental works on fluctuations of injected and
dissipated power in systems with harmonic \cite{Douarche,Joubaud}
and non harmonic potential \cite{Jop,Imparato} within the context
of fluctuations theorem (FT) (see \cite{Kurchan2007} for a review
on FT). The idea of using FT  for theoretically studying the heat
flux in aging systems has been first proposed in
\cite{Crisanti,Zamponi}. Let us summarize the main physical
concepts that are behind this idea. Since non-equilibrium
fluctuating forces due to the collisions of the colloidal glass
particles with a micron-sized probe particle do work $W_{\tau}$
(of ensemble average $\langle W_{\tau} \rangle = 0$) on it in a
given time interval $\tau$, a fraction of $W_{\tau}$ must be
dissipated in the form of heat $Q_{\tau}$. $Q_{\tau}$ is a
stochastic variable which may be either positive (heat received by
the probe) or negative (heat transferred to the colloid). In an
ergodic system acting as an equilibrium thermal bath at
temperature $T$, the mean heat transfer must vanish $\langle
Q_{\tau} \rangle = 0$ in the absence of any external forcing on
the probe in order to satisfy the first law of thermodynamics.
However, in a system out of equilibrium this situation is not
necessarily fulfilled due to non-ergodicity. A mean heat transfer
$\langle Q_{\tau} \rangle > 0$ would be a signature of an
effective temperature of the colloidal glass $T_{eff}(\omega,t_w)
> T$ during a time lag $\tau \approx 2\pi/\omega$. Hence, by investigating
possible asymmetries in the PDF of $Q_{\tau}$ for different time
lags and at different aging times, one could find possible
violations of FT in the aging colloidal glass.

In order to calculate the heat transfer during a time lag $\tau$,
Eq.~(\ref{eq:langvisco}) (with $x_0(t,\omega) = 0$ since we are
interested in the situation without any external forcing) is
multiplied by $\dot{x}$ and integrated from $t_w$ to $t_w+\tau$
(with $\tau \ll t_w$), leading to an extension of the first law of
thermodynamics for a probe particle subject to non-equilibrium
thermal fluctuations of the colloidal glass
\begin{equation}\label{eq:firstlaw}
    \Delta U_{\tau}(t_w) = Q_{\tau}(t_w),
\end{equation}
where
\begin{equation}\label{eq:dU}
    \Delta U_{\tau}(t_w) = \frac{1}{2}(k+K_{Lap}(t_w,\frac{2\pi}{\tau}))(x(t_w+\tau)^2-x(t_w)^2),
\end{equation}
and
\begin{equation}\label{eq:Q}
    Q_{\tau}(t_w) = -\gamma(t_w,\frac{2\pi}{\tau}) \int_{t_w}^{t_w+\tau} \dot{x}(t')^2 dt'+
      \sqrt{2k_B T_{eff}(t_w,\frac{2\pi}{\tau})\gamma(t_w,\frac{2\pi}{\tau})}\int_{t_w}^{t_w+\tau}
      \xi(t')\dot{x}(t')dt'.
\end{equation}
In practice we determine the PDF of the stochastic variable
$Q'_{\tau} \equiv (k/(k+K_{Lap}(t_w,2\pi/\tau)) Q_{\tau}$ by means
of Eqs.~(\ref{eq:firstlaw}) and (\ref{eq:dU}) for particle '2'. At
every aging time $t_w$, we obtain a data set
$\{Q'_{\tau}(t_w)\}_{\tau}$ of $10000-$int$(200$ Hz$\times \tau$) points.
Then we compute the corresponding PDF
at every $t_w$. The results are shown in Fig. \ref{multifig6}. The
PDFs of $Q'_{\tau}$ are exponentially decaying and symmetric with
respect to the maximum value located at $Q'_{\tau} = 0$. At a given
time $t_w$, for different values of $\tau$ we check that there is no
asymmetry in PDFs, which implies that there is no net mean heat flux
taking place at any time scale $\tau$, as shown in Fig.
\ref{multifig6}(a). In addition, Fig. \ref{multifig6}(b) shows that
even during the gelation regime ($t_w > t_g$) no such asymmetries
can be detected. The absence of any asymmetry in the PDF of
$Q_{\tau}$ confirms the absence of any effective temperature of the
colloidal glass higher than the bath temperature even for
fluctuating modes taking place at time scales as slow as $\tau = 5$
s.

\section{Discussion and Conclusion}

We have experimentally studied the statistical properties of the
Brownian motion of a bead inside a colloidal glass of Laponite
during the transition from a liquid to a solid-like state.
Specifically we were interested in understanding whether a
$T_{eff}\ne T$ can be observed  when the Brownian particle is inside
an out of equilibrium bath, i.e. the Laponite suspension in this
specific case. As we have already mentioned, this problem has been
the subject of several measurements giving contrasting results. The
experimental results described in  this paper show not only that,
within experimental errors, $T_{eff}=T$ at any time but they also
indicate  the reasons of the contrasting results. This has been made
possible by the use of multiple optical traps which allow us to
apply simultaneously in the same evolving colloid active and passive
microrheology techniques. Let us summarize the new main results of
this investigation:
\begin{itemize}
\item[a)] As discussed in Sect. \ref{section_sample_preparation},
we have seen that the way the sample is sealed can accelerate the
formation of the gel and the drift of the bead by changing the
chemical properties in the small sample. We have used different
types of cell, Laponite concentrations, bead sizes,
 stiffness  of the optical trap. In each case we do not find any
 increase of the effective temperature.

\item[b)] We propose in Sect. \ref{section_Kramers}
a new method of passive rheology which is based on the use of  the
modulation technique and of the Kramers-Kroening relations. This new
method allows us to estimate $T_{eff}$  and $G$ as a function of
frequency and time. We find that $T_{eff}$ is constant and equal to
$T$ for any frequency and time.

\item[c)] The use of multiple trap allows us to check simultaneously
two microrheology techniques that had led to conflicting results in
the past and to find the possible reasons of this contrast.
 In Sect. \ref{section_properties} we have shown that:
\begin{itemize}
\item[c.1)] In the case of the passive rheology methods based on
the laser modulation the use of too short time windows to measure
the variance of fluctuations and of too long time windows for data
smoothing induces an anomalous increase of the $T_{eff}$ at very
long times which is suppressed if a more precise and unbiased data
analysis is performed. \item[c.2)] By means of AMR at very low
frequencies we check that there is no actual violation of FDT in the
aging colloidal glass. An advantage of the AMR technique is that it
also allows to resolve the slowly evolving viscosity and elasticity
of the colloidal glass in order to observe its transition from a
purely viscous fluid to a viscoelastic one.\item[c.3)] For the
active rheology method we have shown in Fig. \ref{multifig3} the
existence of a maximum as a function of time for the fluctuation
spectra and for the imaginary part of response function, such that
the ratio of these two functions gives a constant $T_{eff}$. The
position of these maxima depends on the strength of the optical
traps used to measure the response and fluctuations. These strengths
are the same in our case whereas are different in the experiment of
Ref. \cite{Abou03}. This difference in the strength induces a shift
on the time position of these maxima which must be corrected in the
data analysis. Small errors in this correction may of course induce
an anomalous time dependence of $T_{eff}$.
 \item[c.4)] The mean position of
the bead during aging evolves. When the stiffness of
   the gel becomes comparable to the optical one,
   the bead starts to move away from the center of the optical trap.
   We observe a drift of the bead position at long time, which could lead
   to the escape of the bead. This effect can be suppressed  by a
   careful preparation of the cell.
\end{itemize}

\item[d)] The use of a new method based on the PDF of heat
fluctuations, Section \ref{section_PDF_heat}. This technique shows
no mean heat transfer between the aging colloidal glass and the
surrounding environment even at long time scales. This is consistent
with the results of passive measurements which show that $T_{eff}$
is equal to $T$ at any time.

\end{itemize}

These results show that the effective temperature of the colloidal
glass and the environment is always the same during aging.
Furthermore they explain where the conflict between the various
results reported in literature may come from. FDT seems to be a very
robust property in colloidal glasses. Our results show no increase
of $T_{eff}$ in Laponite and are in agreement with those of Ref.
\cite{Bellon02_I}, based on macroscopic measurements, and those of
Jabbari-Farouji \emph{et al} \cite{jabbari,jabbari1,jabbari2}, who
measured fluctuations and responses of the bead displacement in
Laponite over a wide frequency range and found that $T_{eff}$ is
equal to the bath temperature. We want to conclude on the fact that
for dielectric measurements in Ref. \cite{Bellon02_I} a very large
$T_{eff}$ has been reported. This is not in contrast with the
results discussed in this paper as $T_{eff}$ is in principle
observable dependent.

\section{Acknowledgement} This work has been partially supported by
ANR-05-BLAN-0105-01.

\end{document}